%% file: Doc.tex
\documentclass[a4paper,UKenglish,cleveref, autoref, thm-restate]{lipics-v2021}

\newtheorem{axiom}{Axiom}

\title{Quorum Subsumption for Heterogeneous Quorum Systems}

 \author{Xiao Li}{University of California, Riverside, USA }{xli289@ucr.edu}{https://orcid.org/0000-0002-1825-0097}{}
 
 \author{Eric Chan}{University of California, Riverside, USA}{echan044@ucr.edu}{[orcid]}{}
 
 \author{Mohsen Lesani}{University of California, Riverside, USA}{lesani@ucr.edu}{[orcid]}{}
\authorrunning{X. Li, E. Chan, and M. Lesani} 

 \Copyright{Xiao Li, Eric Chan, Mohsen Lesani} 

 
\EventEditors{Rotem Oshman}
\EventNoEds{1}
\EventLongTitle{37th International Symposium on Distributed Computing (DISC 2023)}
\EventShortTitle{DISC 2023}
\EventAcronym{DISC}
\EventYear{2023}
\EventDate{October 10-12, 2023}
\EventLocation{L'Aquila, Italy}
\EventLogo{}
\SeriesVolume{281}
\ArticleNo{16}
\input{predefs}

\begin{document}

\nolinenumbers
\tableofcontentsprime{

\ \\
\ \\









☞{Citing the relevant program committee.}

}


\maketitle 
\input{abstract}
\pagestyle{plain}  

\pagenumbering{arabic}




\input{introduction}

\clearpageprime 

\input{quorum_system}
\clearpageprime

\input{ProblemDefinitions}

\clearpageprime

\input{impossibility}

\clearpageprime

\input{ProtocolsEric2}

\clearpageprime

%
%
%
%
%
%
%
%
%









\clearpageprime

\input{related_work}

\input{conclusion}

\section*{Acknowledgments}
We would like to thank DISC '23 reviewers for detailed and constructive reviewers.
Further, we would like to specially thank Giuliano Losa for his insightful comments.

\clearpageprime
\bibliography{Refs}





\end{document}


%% file: predefs.tex
\usepackage{amsmath,amssymb,amsfonts}
\usepackage{graphicx}
\usepackage{textcomp}
\usepackage{xcolor}
\usepackage{mathtools}
\usepackage{algorithmicx}
\usepackage[boxed,ruled,vlined, linesnumbered]{algorithm2e}
\SetVlineSkip{0.3pt}
\SetNlSty{bfseries}{\color{black}}{}
\SetCommentSty{normalfont}
\SetKwComment{Comment}{$\triangleright$\ }{}
\newcommand{\AComment}[1]{\Comment*[r]{#1}}

\newcommand{\nonl}{\renewcommand{\nl}{\let\nl\oldnl}}

\def\BibTeX{{\rm B\kern-.05em{\sc i\kern-.025em b}\kern-.08em
		T\kern-.1667em\lower.7ex\hbox{E}\kern-.125emX}}
\usepackage{tkz-euclide}
\usepackage{mathpartir}
\usepackage{color}
\usepackage{listings}
\usepackage{tikz}
\usetikzlibrary{positioning}
\usepackage{stmaryrd}
\usepackage{comment}
\usepackage{amsthm} 
\usepackage{multirow}
\usepackage{wrapfig}
\usepackage{subcaption}

\newcommand{\angled}[1]{{ \langle #1 \rangle }}

\def\<{\langle}
\def\>{\rangle}
\def\Q{\mathcal{Q}}

\def\P{\mathcal{P}}
\def\B{\mathcal{B}}
\def\W{\mathcal{W}}
\def\F{\mathcal{F}}

\def\Implements{\textbf{Implements}}
\def\Uses{\textbf{Uses}}

\def\colon{\textbf{:}}

\def\request{\textbf{request}}
\def\response{\textbf{response}}

\def\true{\mathsf{true}}
\def\false{\mathsf{false}}

\def\for_all{\textsf{for all}}

\def\candidate{\mathit{candidate}}

\def\apl{\mathit{ptp}}
\def\le{\mathit{le}}

\def\candidate{\mathit{candidate}}
\def\prepared{\mathit{prepared}}

\usepackage{hyperref}
\hypersetup{
    linktoc=all,
    colorlinks=false,       
    linkcolor=black,          
    linkbordercolor=black,
    citecolor=black,        
    filecolor=black,      
    urlcolor=black,           
    pdfborderstyle={/S/U/W 0}
}

\usepackage{tabularx}

\newcommand*\circled[1]{\tikz[baseline=(char.base)]{
		\node[shape=circle,draw,inner sep=2pt] (char) {#1};}}

\usepackage{caption}
\usepackage{graphicx}

\input{Unicode}

\def\ie{\textit{i.e.}}

\def\self{\mathbf{self}}

\definecolor{forestgreen}{rgb}{0.0, 0.27, 0.13}
\definecolor{tropicalrainforest}{rgb}{0.0, 0.46, 0.37}

\newcommand{\tableofcontentsprime}[1]{}

\def\clearpageprime{}
\def\newpageprime{}

\newcommand{\inputprime}[1]{}

\newcommand{\xl}[1]{\textcolor{blue}{{[XL:~#1]}}}

\newcommand{\todo}[1]{\textcolor{tropicalrainforest}{{[Todo:~#1]}}}

\newcommand{\myparagraph}[1]{{\em\medskip\noindent\textbf{#1}.}}
\renewcommand{\paragraph}[1]{\textit{#1. \ }}

\newcommand{\ifspace}[1]{}

\usepackage{multicol}


%% file: Unicode.tex
\DeclareMathAlphabet{\mathpzc}{OT1}{pzc}{m}{it}

\DeclareUnicodeCharacter{261E}{\todo}      

\DeclareUnicodeCharacter{00AC}{\ensuremath{\neg}}                     
\DeclareUnicodeCharacter{00B1}{\ensuremath{\pm}}                      
\DeclareUnicodeCharacter{00B2}{\ensuremath{^2}}                       
\DeclareUnicodeCharacter{00B3}{\ensuremath{^3}}                       
\DeclareUnicodeCharacter{00B7}{\ensuremath{\cdot}}                    
\DeclareUnicodeCharacter{00B9}{\ensuremath{^1}}                       
\DeclareUnicodeCharacter{00D7}{\ensuremath{\times}}                   
\DeclareUnicodeCharacter{02B0}{\ensuremath{^h}}                       
\DeclareUnicodeCharacter{02B3}{\ensuremath{^r}}                       
\DeclareUnicodeCharacter{02E2}{\ensuremath{^s}}                       
\DeclareUnicodeCharacter{20F0}{\ensuremath{^*}}                       
\DeclareUnicodeCharacter{0302}{\ensuremath{\hat{\phantom{x}}}}        
\DeclareUnicodeCharacter{0332}{\ensuremath{\underline{\phantom{x}}}}  
\DeclareUnicodeCharacter{0308}{\ensuremath{\ddot{\phantom{x}}}}       
\DeclareUnicodeCharacter{0393}{\ensuremath{\Gamma}}                   
\DeclareUnicodeCharacter{0394}{\ensuremath{\Delta}}                   
\DeclareUnicodeCharacter{0398}{\ensuremath{\Theta}}                   
\DeclareUnicodeCharacter{039B}{\ensuremath{\Lambda}}                  
\DeclareUnicodeCharacter{039E}{\ensuremath{\Xi}}                      
\DeclareUnicodeCharacter{03A0}{\ensuremath{\Pi}}                      
\DeclareUnicodeCharacter{03A3}{\ensuremath{\Sigma}}                   

\DeclareUnicodeCharacter{03A4}{\ensuremath{\mathrm{T}}}  

\DeclareUnicodeCharacter{03A5}{\ensuremath{\Upsilon}}   
\DeclareUnicodeCharacter{03A6}{\ensuremath{\Phi}}       
\DeclareUnicodeCharacter{03A8}{\ensuremath{\Psi}}       
\DeclareUnicodeCharacter{03A9}{\ensuremath{\Omega}}     
\DeclareUnicodeCharacter{03B1}{\ensuremath{\alpha}}     
\DeclareUnicodeCharacter{03B2}{\ensuremath{\beta}}      
\DeclareUnicodeCharacter{03B3}{\ensuremath{\gamma}}     
\DeclareUnicodeCharacter{03B4}{\ensuremath{\delta}}     
\DeclareUnicodeCharacter{03B5}{\ensuremath{\epsilon}}   
\DeclareUnicodeCharacter{03B6}{\ensuremath{\zeta}}      
\DeclareUnicodeCharacter{03B7}{\ensuremath{\eta}}       
\DeclareUnicodeCharacter{03B8}{\ensuremath{\theta}}     
\DeclareUnicodeCharacter{03B9}{\ensuremath{\iota}}      
\DeclareUnicodeCharacter{03BA}{\ensuremath{\kappa}}     
\DeclareUnicodeCharacter{03BB}{\ensuremath{\lambda}}    
\DeclareUnicodeCharacter{03BC}{\ensuremath{\mu}}        
\DeclareUnicodeCharacter{03BD}{\ensuremath{\nu}}        
\DeclareUnicodeCharacter{03BE}{\ensuremath{\xi}}        
\DeclareUnicodeCharacter{03C0}{\ensuremath{\pi}}        
\DeclareUnicodeCharacter{03C1}{\ensuremath{\rho}}       
\DeclareUnicodeCharacter{03C2}{\ensuremath{\varsigma}}  
\DeclareUnicodeCharacter{03C3}{\ensuremath{\sigma}}     
\DeclareUnicodeCharacter{03C4}{\ensuremath{\tau}}       
\DeclareUnicodeCharacter{03C5}{\ensuremath{\upsilon}}   
\DeclareUnicodeCharacter{03C6}{\ensuremath{\varphi}}    
\DeclareUnicodeCharacter{03C7}{\ensuremath{\chi}}       
\DeclareUnicodeCharacter{03C8}{\ensuremath{\psi}}       
\DeclareUnicodeCharacter{03C9}{\ensuremath{\omega}}     
\DeclareUnicodeCharacter{03D1}{\ensuremath{\vartheta}}  

\DeclareUnicodeCharacter{0432}{\ensuremath{\mathrm{PDF\;TODO}}}  
\DeclareUnicodeCharacter{0435}{\ensuremath{\mathrm{PDF\;TODO}}}  
\DeclareUnicodeCharacter{0438}{\ensuremath{\mathrm{PDF\;TODO}}}  
\DeclareUnicodeCharacter{043C}{\ensuremath{\mathrm{PDF\;TODO}}}  
\DeclareUnicodeCharacter{0440}{\ensuremath{\mathrm{PDF\;TODO}}}  
\DeclareUnicodeCharacter{0442}{\ensuremath{\mathrm{PDF\;TODO}}}  
\DeclareUnicodeCharacter{041F}{\ensuremath{\mathrm{PDF\;TODO}}}  
\DeclareUnicodeCharacter{0BA8}{\ensuremath{\mathrm{PDF\;TODO}}}  
\DeclareUnicodeCharacter{0BBF}{\ensuremath{\mathrm{PDF\;TODO}}}  
\DeclareUnicodeCharacter{1100}{\ensuremath{\mathrm{PDF\;TODO}}}  
\DeclareUnicodeCharacter{11F9}{\ensuremath{\mathrm{PDF\;TODO}}}  

\DeclareUnicodeCharacter{1D48}{\ensuremath{^d}}  
\DeclareUnicodeCharacter{1D50}{\ensuremath{^m}}  
\DeclareUnicodeCharacter{1D56}{\ensuremath{^p}}  
\DeclareUnicodeCharacter{1D62}{\ensuremath{_i}}  
\DeclareUnicodeCharacter{1D63}{\ensuremath{_r}}  
\DeclareUnicodeCharacter{1D64}{\ensuremath{_u}}  
\DeclareUnicodeCharacter{1D9C}{\ensuremath{^c}}  

\DeclareUnicodeCharacter{2032}{\ensuremath{\text{\textasciiacute}}}  

\DeclareUnicodeCharacter{2070}{\ensuremath{^0}}  
\DeclareUnicodeCharacter{2074}{\ensuremath{^4}}  
\DeclareUnicodeCharacter{2075}{\ensuremath{^5}}  
\DeclareUnicodeCharacter{2076}{\ensuremath{^6}}  
\DeclareUnicodeCharacter{2077}{\ensuremath{^7}}  
\DeclareUnicodeCharacter{2078}{\ensuremath{^8}}  
\DeclareUnicodeCharacter{2079}{\ensuremath{^9}}  
\DeclareUnicodeCharacter{207A}{\ensuremath{^+}}  
\DeclareUnicodeCharacter{207B}{\ensuremath{^-}}  
\DeclareUnicodeCharacter{207F}{\ensuremath{^n}}  
\DeclareUnicodeCharacter{2080}{\ensuremath{_0}}  
\DeclareUnicodeCharacter{2081}{\ensuremath{_1}}  
\DeclareUnicodeCharacter{2082}{\ensuremath{_2}}  
\DeclareUnicodeCharacter{2083}{\ensuremath{_3}}  
\DeclareUnicodeCharacter{2084}{\ensuremath{_4}}  
\DeclareUnicodeCharacter{2085}{\ensuremath{_5}}  
\DeclareUnicodeCharacter{2086}{\ensuremath{_6}}  
\DeclareUnicodeCharacter{2087}{\ensuremath{_7}}  
\DeclareUnicodeCharacter{2088}{\ensuremath{_8}}  
\DeclareUnicodeCharacter{2089}{\ensuremath{_9}}  
\DeclareUnicodeCharacter{208A}{\ensuremath{_+}}  
\DeclareUnicodeCharacter{208B}{\ensuremath{_-}}  
\DeclareUnicodeCharacter{2091}{\ensuremath{_e}}  
\DeclareUnicodeCharacter{2095}{\ensuremath{_h}}  
\DeclareUnicodeCharacter{2096}{\ensuremath{_k}}  
\DeclareUnicodeCharacter{2097}{\ensuremath{_l}}  
\DeclareUnicodeCharacter{2098}{\ensuremath{_m}}  
\DeclareUnicodeCharacter{2099}{\ensuremath{_n}}  
\DeclareUnicodeCharacter{209A}{\ensuremath{_p}}  
\DeclareUnicodeCharacter{2092}{\ensuremath{_o}}  
\DeclareUnicodeCharacter{209B}{\ensuremath{_s}}  
\DeclareUnicodeCharacter{209C}{\ensuremath{_t}}  

\DeclareUnicodeCharacter{2113}{\ensuremath{\ell}} 

\DeclareUnicodeCharacter{2115}{\ensuremath{\mathbbm{N}}}  
\DeclareUnicodeCharacter{211A}{\ensuremath{\mathbbm{Q}}}  
\DeclareUnicodeCharacter{211D}{\ensuremath{\mathbbm{R}}}  
\DeclareUnicodeCharacter{2124}{\ensuremath{\mathbbm{Z}}}  

\DeclareUnicodeCharacter{2135}{\ensuremath{\aleph}}           
\DeclareUnicodeCharacter{2190}{\ensuremath{\leftarrow}}       
\DeclareUnicodeCharacter{2192}{\ensuremath{\rightarrow}}      
\DeclareUnicodeCharacter{2194}{\ensuremath{\leftrightarrow}}  

\DeclareUnicodeCharacter{21A0}{\ensuremath{\twoheadrightarrow}}  

\DeclareUnicodeCharacter{21A6}{\ensuremath{\mapsto}}  
\DeclareUnicodeCharacter{21A6}{\ensuremath{\mapsto}}  

\DeclareUnicodeCharacter{21BE}{\ensuremath{\restriction}}  

\DeclareUnicodeCharacter{21D0}{\ensuremath{\Leftarrow}}       
\DeclareUnicodeCharacter{21D2}{\ensuremath{\Rightarrow}}      
\DeclareUnicodeCharacter{21D4}{\ensuremath{\Leftrightarrow}}  

\DeclareUnicodeCharacter{2200}{\ensuremath{\forall}}      
\DeclareUnicodeCharacter{2203}{\ensuremath{\exists}}      
\DeclareUnicodeCharacter{2204}{\ensuremath{\not\exists}}  
\DeclareUnicodeCharacter{2205}{\ensuremath{\emptyset}}    
\DeclareUnicodeCharacter{2208}{\ensuremath{\in}}          
\DeclareUnicodeCharacter{2209}{\ensuremath{\not\in}}      
\DeclareUnicodeCharacter{220B}{\ensuremath{\ni}}          
\DeclareUnicodeCharacter{220C}{\ensuremath{\not\ni}}      

\DeclareUnicodeCharacter{220E}{{\tiny \ensuremath{\blacksquare}}} 

\DeclareUnicodeCharacter{220F}{{\tiny \ensuremath{\prod}}}  
\DeclareUnicodeCharacter{2211}{\ensuremath{\sum}}           
\DeclareUnicodeCharacter{2218}{\ensuremath{\circ}}          
\DeclareUnicodeCharacter{2219}{\ensuremath{\bullet}}        
\DeclareUnicodeCharacter{221E}{\ensuremath{\infty}}         
\DeclareUnicodeCharacter{2223}{\ensuremath{\mid}}           
\DeclareUnicodeCharacter{2225}{\ensuremath{\parallel}}      
\DeclareUnicodeCharacter{2227}{\ensuremath{\wedge}}         
\DeclareUnicodeCharacter{2228}{\ensuremath{\vee}}           

\DeclareUnicodeCharacter{2229}{\ensuremath{\cap}}           

\DeclareUnicodeCharacter{222A}{\ensuremath{\cup}}           
\DeclareUnicodeCharacter{222B}{\ensuremath{\int}}           
\DeclareUnicodeCharacter{2234}{\ensuremath{\therefore}}     

\DeclareUnicodeCharacter{2237}{\ensuremath{\dblcolon}} 

\newcommand{\dotminus}{\mathop{\mbox{$-^{\hspace{-.5em}\cdot}\,$}}}
\DeclareUnicodeCharacter{2238}{\ensuremath{\dotminus}}  

\DeclareUnicodeCharacter{223C}{\ensuremath{\sim}}       
\DeclareUnicodeCharacter{2243}{\ensuremath{\simeq}}     
\DeclareUnicodeCharacter{2245}{\ensuremath{\cong}}      
\DeclareUnicodeCharacter{2248}{\ensuremath{\approx}}    
\DeclareUnicodeCharacter{2250}{\ensuremath{\doteq}}     

\DeclareUnicodeCharacter{2254}{\ensuremath{\coloneqq}}  

\newcommand{\eqq}{\stackrel{\text{\tiny ?}}{=}}
\DeclareUnicodeCharacter{225F}{\ensuremath{\eqq}}  

\DeclareUnicodeCharacter{2260}{\ensuremath{\neq}}         
\DeclareUnicodeCharacter{2261}{\ensuremath{\equiv}}       
\DeclareUnicodeCharacter{2262}{\ensuremath{\not\equiv}}   
\DeclareUnicodeCharacter{2264}{\ensuremath{\le}}          
\DeclareUnicodeCharacter{2265}{\ensuremath{\ge}}          
\DeclareUnicodeCharacter{226E}{\ensuremath{\nless}}       
\DeclareUnicodeCharacter{226F}{\ensuremath{\ngtr}}        
\DeclareUnicodeCharacter{2270}{\ensuremath{\nleq}}        
\DeclareUnicodeCharacter{2271}{\ensuremath{\ngeq}}        
\DeclareUnicodeCharacter{227A}{\ensuremath{\prec}}        
\DeclareUnicodeCharacter{2280}{\ensuremath{\nprec}}        
\DeclareUnicodeCharacter{227C}{\ensuremath{\preceq}}      
\DeclareUnicodeCharacter{2282}{\ensuremath{\subset}}      
\DeclareUnicodeCharacter{2284}{\ensuremath{\not\subset}} 
\DeclareUnicodeCharacter{2283}{\ensuremath{\supset}}      
\DeclareUnicodeCharacter{2286}{\ensuremath{\subseteq}}    
\DeclareUnicodeCharacter{228E}{\ensuremath{\uplus}}       
\DeclareUnicodeCharacter{2291}{\ensuremath{\sqsubseteq}}  
\DeclareUnicodeCharacter{2293}{\ensuremath{\sqcap}}       
\DeclareUnicodeCharacter{2294}{\ensuremath{\sqcup}}       
\DeclareUnicodeCharacter{2295}{\ensuremath{\oplus}}       
\DeclareUnicodeCharacter{22A2}{\ensuremath{\vdash}}       
\DeclareUnicodeCharacter{22A4}{\ensuremath{\top}}         
\DeclareUnicodeCharacter{22A5}{\ensuremath{\bot}}         
\DeclareUnicodeCharacter{22C0}{\ensuremath{\bigwedge}}    
\DeclareUnicodeCharacter{22C2}{\ensuremath{\bigcap}}      
\DeclareUnicodeCharacter{22C3}{\ensuremath{\bigcup}}      
\DeclareUnicodeCharacter{22EE}{\ensuremath{\vdots}}       
\DeclareUnicodeCharacter{22EF}{\ensuremath{\cdots}}       

\DeclareUnicodeCharacter{22F0}{\ensuremath{\iddots}} 

\DeclareUnicodeCharacter{230A}{\ensuremath{\lfloor}}  
\DeclareUnicodeCharacter{230B}{\ensuremath{\rfloor}}  
\DeclareUnicodeCharacter{2308}{\ensuremath{\lceil}}   
\DeclareUnicodeCharacter{2309}{\ensuremath{\rceil}}   
\DeclareUnicodeCharacter{2329}{\ensuremath{\langle}}  
\DeclareUnicodeCharacter{232A}{\ensuremath{\rangle}}  

\DeclareUnicodeCharacter{23CE}{\ensuremath{\Return}}  

\DeclareUnicodeCharacter{2460}{\ensuremath{\text{\ding{172}}}} 
\DeclareUnicodeCharacter{2461}{\ensuremath{\text{\ding{173}}}} 
\DeclareUnicodeCharacter{2462}{\ensuremath{\text{\ding{174}}}} 
\DeclareUnicodeCharacter{2463}{\ensuremath{\text{\ding{175}}}} 
\DeclareUnicodeCharacter{2464}{\ensuremath{\text{\ding{176}}}} 
\DeclareUnicodeCharacter{2465}{\ensuremath{\text{\ding{177}}}} 
\DeclareUnicodeCharacter{2466}{\ensuremath{\text{\ding{178}}}} 
\DeclareUnicodeCharacter{2467}{\ensuremath{\text{\ding{179}}}} 
\DeclareUnicodeCharacter{2468}{\ensuremath{\text{\ding{180}}}} 

\DeclareUnicodeCharacter{25C5}{\ensuremath{\triangleleft}}  

\DeclareUnicodeCharacter{2660}{\ensuremath{\spadesuit}} 
\DeclareUnicodeCharacter{266D}{\ensuremath{\flat}}      
\DeclareUnicodeCharacter{266F}{\ensuremath{\sharp}}     

\DeclareUnicodeCharacter{2713}{\ensuremath{\checkmark}}  

\DeclareUnicodeCharacter{27E6}{\ensuremath{\llbracket}}  
\DeclareUnicodeCharacter{27E7}{\ensuremath{\rrbracket}}  

\DeclareUnicodeCharacter{27E8}{\ensuremath{\langle}}          
\DeclareUnicodeCharacter{27E9}{\ensuremath{\rangle}}          
\DeclareUnicodeCharacter{27F6}{\ensuremath{\longrightarrow}}  

\DeclareUnicodeCharacter{2983}{\ensuremath{\mathrm{PDF\;TODO}}}  
\DeclareUnicodeCharacter{2984}{\ensuremath{\mathrm{PDF\;TODO}}}  
\DeclareUnicodeCharacter{2985}{\ensuremath{\mathrm{PDF\;TODO}}}  
\DeclareUnicodeCharacter{2986}{\ensuremath{\mathrm{PDF\;TODO}}}  

\DeclareUnicodeCharacter{2987}{\ensuremath{\llparenthesis}}  
\DeclareUnicodeCharacter{2988}{\ensuremath{\rrparenthesis}}  

\DeclareUnicodeCharacter{2216}{\ensuremath{\setminus}}

\DeclareUnicodeCharacter{2A06}{\ensuremath{\bigcup}}  
\DeclareUnicodeCharacter{2C7C}{\ensuremath{_j}}       

\DeclareUnicodeCharacter{D7B0}{\ensuremath{\mathrm{PDF\;TODO}}} 

\DeclareUnicodeCharacter{1D7D9}{\ensuremath{\mathbbm{1}}}  

\DeclareUnicodeCharacter{1D4D0}{\ensuremath{\mathcal{A}}} 
\DeclareUnicodeCharacter{1D4D1}{\ensuremath{\mathcal{B}}}  
\DeclareUnicodeCharacter{1D4D2}{\ensuremath{\mathcal{C}}}  
\DeclareUnicodeCharacter{1D4D3}{\ensuremath{\mathcal{D}}}  
\DeclareUnicodeCharacter{1D4D4}{\ensuremath{\mathcal{E}}}  
\DeclareUnicodeCharacter{1D4D5}{\ensuremath{\mathcal{F}}}  
\DeclareUnicodeCharacter{1D4D6}{\ensuremath{\mathcal{G}}}  
\DeclareUnicodeCharacter{1D4D7}{\ensuremath{\mathcal{H}}}  
\DeclareUnicodeCharacter{1D4D8}{\ensuremath{\mathcal{I}}}  
\DeclareUnicodeCharacter{1D4D9}{\ensuremath{\mathcal{J}}}  
\DeclareUnicodeCharacter{1D4DA}{\ensuremath{\mathcal{K}}}  
\DeclareUnicodeCharacter{1D4DB}{\ensuremath{\mathcal{L}}}  
\DeclareUnicodeCharacter{1D4DC}{\ensuremath{\mathcal{M}}}  
\DeclareUnicodeCharacter{1D4DD}{\ensuremath{\mathcal{N}}}  
\DeclareUnicodeCharacter{1D4DE}{\ensuremath{\mathcal{O}}}  
\DeclareUnicodeCharacter{1D4DF}{\ensuremath{\mathcal{P}}}

\DeclareUnicodeCharacter{1D4E0}{\ensuremath{\mathcal{Q}}}  
\DeclareUnicodeCharacter{1D4E1}{\ensuremath{\mathcal{R}}}  
\DeclareUnicodeCharacter{1D4E2}{\ensuremath{\mathcal{S}}}  
\DeclareUnicodeCharacter{1D4E3}{\ensuremath{\mathcal{T}}}  
\DeclareUnicodeCharacter{1D4E4}{\ensuremath{\mathcal{U}}}  
\DeclareUnicodeCharacter{1D4E5}{\ensuremath{\mathcal{V}}}  
\DeclareUnicodeCharacter{1D4E6}{\ensuremath{\mathcal{W}}}  
\DeclareUnicodeCharacter{1D4E7}{\ensuremath{\mathcal{X}}}  
\DeclareUnicodeCharacter{1D4E8}{\ensuremath{\mathcal{Y}}}  
\DeclareUnicodeCharacter{1D4E9}{\ensuremath{\mathcal{Z}}}  

\DeclareUnicodeCharacter{25C2}{\ensuremath{\triangleleft}}  
\DeclareUnicodeCharacter{25B9}{\ensuremath{\triangleright}}
\DeclareUnicodeCharacter{2286}{\ensuremath{\subseteq}}    
\DeclareUnicodeCharacter{2217}{\em}

\DeclareUnicodeCharacter{1D62}{\ensuremath{_i}}  
\DeclareUnicodeCharacter{2C7C}{\ensuremath{_j}}
\DeclareUnicodeCharacter{2096}{\ensuremath{_k}}

\DeclareUnicodeCharacter{276A}{\mbox{$($}}  
\DeclareUnicodeCharacter{276B}{\mbox{$)$}}  

\DeclareUnicodeCharacter{261E}{\todo}      

\DeclareUnicodeCharacter{1D539}{\mathbb{B}}      
\DeclareUnicodeCharacter{2102}{\mathbb{C}}      
\DeclareUnicodeCharacter{2098}{_m}      
\DeclareUnicodeCharacter{2115}{\mathbb{N}}      

\DeclareUnicodeCharacter{22E6}{\mathbb{\; \lnsim \; }}      

%% file: abstract.tex
\begin{abstract}
Byzantine quorum systems 
provide higher throughput than proof-of-work
and
incur modest energy consumption.
Further,
their modern incarnations
incorporate
personalized and heterogeneous trust.
Thus, they are emerging as an appealing candidate for
global financial infrastructure.
However, since their quorums are not uniform across processes anymore,
the properties that they should maintain
to support abstractions such as reliable broadcast and consensus
are not well-understood.
It has been shown that the two properties
quorum intersection and availability are necessary.
In this paper, we prove that they are not sufficient.
We then define the 
notion of quorum subsumption,
and
show that the three conditions together are sufficient:
we 
present
reliable broadcast and consensus protocols,
and prove their correctness 
for quorum systems that provide the three properties.
\end{abstract}

%% file: introduction.tex
\section{Introduction}


Bitcoin \cite{nakamoto2008peer} had the promise to democratize the global finance.
Globally scattered servers validate and process transactions,
and maintain a consistent replication of a ledger.
However, 
the nature of
the proof-of-work consensus 
exhibited disadvantages such as
high energy consumption,
and
low throughput.
In contrast, 
Byzantine replication have always had modest energy consumption.
Further,
since its advent as PBFT \cite{castro1999practical},
many recent extensions 
\cite{veronese2011efficient, miller2016honey, yin2019hotstuff, carr2022towards, baudet2019state, buchman2016tendermint,buchman2022revisiting}
have improved its throughput.
However, 
its basic model of quorums is closed and homogeneous:
the set of processes are fixed,
and
the quorums are assumed to be uniform across processes.
Thus, projects such as 
Ripple \cite{schwartz2014ripple} and Stellar \cite{mazieres2015stellar,lokhava2019fast}
emerged to bring 
{∗heterogeneity}
and openness
to Byzantine quorum systems.
%
They let every process declare its own set of quorums,
or the processes it trusts called slices,
from which quorums are calculated.


In this paper, we first consider 
a basic model of heterogeneous quorum systems
where each process has an individual set of quorums.
Then, we consider fundamental questions about their properties.
Quorum systems are the foundation of common distributed computing abstractions such as reliable broadcast and consensus.
We specify the expected safety and liveness properties for these abstractions.
{∗What are the necessary and sufficient properties of heterogeneous quorum systems
to support these abstractions?}
%
Previous work \cite{losa2019stellar} noted that 
quorum intersection and weak availability properties are necessary
for the quorum system to implement the consensus abstraction.
Quorum intersection requires that 
every pair of quorums 
overlap at a well-behaved process.
%
The safety of consensus relies on 
the quorum intersection
property of the underlying quorum system:
intuitively, if an operation communicates with a quorum, and 
a later operation communicates with another quorum, 
a single well-behaved process in their intersection 
can make the second quorum aware of 
the first.
%
%
A quorum system is 
weakly
available for a process
if it has a quorum for that process whose members are all well-behaved.
Intuitively,
the quorum system is available
to that process through that quorum.
Since a process needs to communicate with at least one quorum to terminate,
the liveness properties 
are dependent on
the 
availability of the quorum system.

The quorum intersection and availability properties are {∗necessary}.
Are they 
sufficient as well?
In this paper, we prove that they are {∗not sufficient} conditions to implement reliable broadcast and consensus.
For each abstraction, 
we present execution scenarios,
and apply indistinguishability arguments
to show that 
any protocol violates at least one of the safety or liveness properties.
What property should be added to make the properties sufficient?
A less known property 
is quorum sharing \cite{losa2019stellar}.
Roughly speaking, 
every quorum should include a quorum for all its members.
This is a property that trivially holds for homogeneous quorum systems 
where every quorum is uniformly a quorum of all its members.
However, in general, it does not hold for heterogeneous quorum systems.
Previous work showed that 
it also holds 
for Stellar quorums
if Byzantine processes do not lie about their slices.

Since Byzantine processes' quorums is arbitrary,
in practice, quorum sharing is too strong.
In order to require inclusion only for the quorums of a well-behaved subset of processes,
we consider
a weaker notion, called {∗quorum subsumption}.
As we will see, this property
lets processes in the included quorum make local decisions
while preserving the properties of the including quorum.
We 
precisely
capture this property, and
show that together with the other two properties,
it is sufficient to implement
reliable broadcast and consensus abstractions.
We 
present
{∗protocols for both reliable broadcast and consensus},
and prove that 
if the underlying quorum system has quorum intersection,
availability,
and 
subsumption for certain quorums,
then the protocols satisfy the required safety and liveness properties.
%
%

In summary, this paper makes the following contributions.
\begin{itemize}
%
        \item
        Properties of quorum-based protocols (\autoref{sec:proto-impl})
        and 
        specifications of reliable broadcast and consensus 
        on heterogeneous quorum systems (\autoref{sec:protocol-spec}).

        \item
        Proof of insufficiency of quorum intersection and availability to solve consensus (\autoref{sec:consensus-imp}) and reliable broadcast (\autoref{sec:reliable-broad-imp}).        
        
        
        \item 
        Sufficiency of 
        quorum intersection, quorum availability and quorum subsumption to solve consensus and reliable broadcast.
        We present protocols for 
        reliable broadcast (\autoref{sec:proto-reliable-broad}) and 
        consensus (\autoref{sec:proto-consensus}),
        and their proofs of correctness.
        

\end{itemize}


%% file: quorum_system.tex
\section{Heterogeneous Quorum Systems}
\label{sec:q-systems}

A quorum is a subset of processes that are collectively trusted to perform an operation.
However, this trust may not be uniform: 
while a process may trust a part of a system,
another process may not trust that same part.
%
In this section, 
we adopt a general model of 
quorum systems \cite{li2023open,losa2019stellar} and its properties.
These basic definitions adapt common properties of quorum systems to the heterogeneous setting,
and 
serve as the foundation for theorems and protocols in the later sections.
Since we want the theorems to be as strong as possible,
we introduce the weak notion of quorum subsumption
in this paper.

\subsection{Processes and Quorums}



\myparagraph{Processes and Failures}
A quorum system is hosted on a set of processes $𝓟$.
%
For every execution, we can partition
the set 
$𝓟$
into
\emph{Byzantine} $𝓑$
and
\emph{well-behaved} $𝓦 = 𝓟 ∖ 𝓑$
processes.
Well-behaved processes follow the given protocol, while Byzantine processes can deviate from the protocol arbitrarily.


We assume that the network is partially synchronized, 
\ie, after an unknown global stabilization time (GST), 
if both the sender and receiver are well-behaved, the message will eventually be delivered with a known bounded delay \cite{dwork1988consensus}.

\myparagraph{Heterogeneous Quorum Systems (HQS)}
To represent subjective trust, we let each process specify its own quorums. 
A quorum $q$ of process $p$ is a non-empty subset $P$ of $𝓟$ that $p$ trusts 
to get information from if it obtains the same information from each member of $P$.
(In practice, a quorum of $p$ can contain $p$ itself,
although the model does not require it.)
Each process $p$ stores its own set of quorums
that we call individual quorums of $p$. 
Any superset of a quorum of $p$ is also a quorum of $p$;
thus, there are minimal quorums:
a quorum of $p$ is a minimal quorum of $p$ if none of its strict subsets is a quorum of $p$.
Thus, to avoid redundancy, 
$p$ can ignore its quorums that are proper supersets of its minimal quorums.
Thus, each process stores only its individual minimal quorums.


\begin{definition}[Quorum System]
    A heterogeneous quorum system $𝓠$ is 
    a mapping from 
    processes 
    to their
    non-empty
    set of individual minimal quorums.
\end{definition}


%
Since the trust assumptions of Byzantine processes can be arbitrary,
their quorums can be left unspecified.
\autoref{fig:running-example} presents an example quorum system.
%
%
When obvious from the context, 
we say quorums of $p$ to refer to the individual minimal quorums of $p$,
and
use $\Q$ to refer to the set of all individual minimal quorums of the system, i.e. the co-domain of $\Q$.
Additionally, we say quorum systems to refer to heterogeneous quorum systems.
A process $p$ is a \emph{follower} of a process $p'$ iff there is a quorum $q ∈ 𝓠(p)$
that includes $p'$.

In dissemination quorum system (DQS) \cite{malkhi1998byzantine}
(and the cardinality-based quorum systems as a special case),
quorums are uniform for all processes.
%
Processes have the same set of individual minimal quorums.
For example, a quorum system that tolerates $f$ Byzantine failures out of $3f + 1$ processes considers any set of $2f + 1$ processes as a quorum for all processes.

\input{properties}

%% file: properties.tex
\subsection{Properties}


A quorum system is expected to maintain certain properties in order to provide distributed 
abstractions
such as Byzantine reliable broadcast and consensus.
%
Quorum intersection and quorum availability are well-established requirements for quorum systems.
In the following section,
we will see their adaption to HQS.
Further, we identify a new property we call quorum subsumption that helps achieve the aforementioned abstractions on HQS.
%
Finally, we briefly present a few related quorum systems, and their properties.


\label{sec:prop-q-sys}


\myparagraph{Quorum Intersection}
Processes store and retrieve information from the quorum system by communicating with its quorums.
To ensure that
information is properly passed from a quorum to another, 
the quorum system is expected to maintain a well-behaved process at the intersection of every pair of quorums.
For example, in the running example in \autoref{fig:running-example}, all the quorums of well-behaved processes intersect at at least one of well-behaved processes in $\{1, 3, 4\}$.

%


%
%
%

\begin{definition}[Quorum Intersection]\label{def:q-intersect}
A quorum system $\Q$ has quorum intersection 
iff every pair of quorums of well-behaved processes in $\Q$ intersect at a well-behaved process,
\ie,
$\forall p, p' \in \W. \ q \in \Q(p). \ q' \in \Q(p'). \ q \cap q' \cap 𝓦 ≠ ∅$
\end{definition}

\myparagraph{Quorum Availability}
In order to support progress for a process, 
the quorum system is expected to have at least one quorum 
for that process whose members are all well-behaved.
We say that the quorum system is weakly available for that process.
(In the literature, this notion of availability is often unqualified, but we explicitly contrast the weak notion to the strong notion that we will define.)
In classical quorum systems, any quorum is a quorum for all processes.
This guarantees that if 
the quorum system is available for a process,
it is available for all processes.
However, this is obviously not true in a heterogeneous quorum system
where quorums are not uniform.
In this setting, 
we weaken the availability property
so that it requires
only a subset and not necessarily all well-behaved processes to have a well-behaved quorum.
In \autoref{fig:running-example}, $\Q$ is available for the set $\{1, 3, 4\}$: 
the quorum $\{1, 4\}$ of process $1$, and the quorum $\{3, 4\}$ of processes $3$ and $4$ make them weakly available.
Each process in that subset
can always communicate with a quorum
independently of Byzantine processes.


\begin{definition}[Weak Availability]
    \label{def:weak-available-set}
    \label{def:QS-availability}
    A quorum system 
    is weakly available
    for a set of processes $P$
    iff
    every process in $P$ has at least one quorum that is a subset of 
    well-behaved processes $𝓦$.    
    A quorum system is available iff it is available for a non-empty set of processes.
\end{definition}

    


If a quorum system is weakly available,
there is at least one well-behaved process that can communicate with a quorum
independently of Byzantine processes.

With quorum availability introduced, we can consider when a quorum system is unavailable.
A quorum system is unavailable for a process when that process has no quorum in $\W$,
\ie, the Byzantine processes $𝓑$ can block every one of its quorums.
We generalize this idea in the notion of blocking.

\begin{wrapfigure}{R}{0.4\textwidth}
\small
$
\begin{array}{l}
𝓟 = 𝓦 ∪ 𝓑, \ \ 𝓦 = \{ 1 , 3, 4, 5 \}, \ \ 𝓑 = \{ 2 \}
\\
𝓠 = \{ 1 ↦ \{ \{ 1, 2, 3 \}, \{1, 4 \} \},
\\ \phantom{𝓠 = \{ }
           3 ↦ \{ \{ 3, 4 \}, \{1, 3\}\}
\\ \phantom{𝓠 = \{ }
		   4 ↦ \{ \{ 3, 4 \}\}
\\ \phantom{𝓠 = \{ }
			5 ↦ \{\{1, 2, 3, 5\}\}
\}
\end{array}
$
\caption{Quorum System 
Example}
\label{fig:running-example}
\end{wrapfigure}

\begin{definition}[Blocking Set]\label{def:blocking-set}
A set of processes $P$ is a blocking set for a process $p$ (or is $p$-blocking) 
if $P$ intersects every quorum of $p$.
\end{definition}

For example, consider cardinality-based quorum systems where
the system contains $3f+1$ processes.
Any set of size $f+1$ is a blocking set for all well-behaved processes,
since 
a set with $f+1$ processes intersects with any quorum, a set with $2f+1$ processes.
In \autoref{fig:running-example}, 
well-behaved process $5$ is blocked by $\{2\}$, 
since its only quorum $\{1, 2, 3, 5\}$ intersect with 
$\{2\}$

Notice also that the definition does not stipulate that the blocking set is Byzantine, but rather it is more general.
The concept of blocking will be useful for designing our protocols in (\autoref{sec:proto}).
For now, we prove a lemma for blocking sets.
In order to state the lemma, 
we generalize the notion of availability.
Given a set of processes $P$,
we generalize availability for $P$ at the complete set of well-behaved processes $𝓦$
(\autoref{def:weak-available-set})
to availability for $P$ at a subset $P'$ of well-behaved processes.
%
We say that a quorum system 
is weakly available
for a set of processes $P$
at a subset of well-behaved processes $P'$
iff
every process in $P$ has at least one quorum that is a subset of $P'$.


\begin{lemma}
	\label{lem:block-w-available}
	In every 
	quorum system
	that is
	weakly available for a set of processes $P$ at $P'$,
	every blocking set of
	every process in $P$
	intersects $P'$.
\end{lemma}

\begin{proof}
	Consider a quorum system 
	that is weakly available for $P$ at $P'$,
	a process $p$ in $P$,
	and a set of processes $P''$ that blocks $p$.
	By the definition of available,
	there is at least one quorum $q$ of $p$ that is a subset of $P'$.
	By the definition of blocking set (\autoref{def:blocking-set}),
	$q$ intersects with $P''$.
	Hence, $P'$ intersects $P''$
	as well.
\end{proof}

\myparagraph{Quorum subsumption}
%
We now introduce the notion of quorum subsumption.


%

\begin{definition}[Quorum Subsumption]
    \label{def:quorum-subsumption}
    A quorum system $\Q$ is quorum subsuming for a quorum $q$
    iff
    every process in $q$
    has a quorum 
    that is included in $q$,    
%
%
    \ie,
    $∀p ∈ q. \ 
    ∃q' ∈ 𝓠(p). \ 
    q' ⊆ q$. 
    We say that $𝓠$ is quorum subsuming for a set of quorums 
    if it is quorum subsuming for each quorum in the set.
\end{definition}

In \autoref{fig:running-example}, 
$\Q$ is quorum subsuming for $\{ 3, 4 \}$: both members in this quorum have the quorum $\{3, 4\}$ that is trivially a subset of itself. 
However, $\Q$ is not quorum subsuming for process $1$'s quorum $\{1, 4\}$: 
process $4$'s only quorum $\{3, 4\}$ is not a subset of $\{1, 4\}$.

\begin{wraptable}[8]{R}{0.7\textwidth}
	\small
	\centering
	\begin{tabular}{|c @{} c @{} c @{} c @{} c|} 
		\hline
		sender & 1 & \textcolor{red}{2} & 3 & 4 \\ 
		\hline
		$\mathit{BCast(m_1)}$ &  &  &  & \\ 
		\hline
		& $\mathit{Echo(m_1)}$ &  & $\mathit{Echo(m_1)}$ & $\mathit{Echo(m_1)}$ \\
		\hline
		&  & \textcolor{red}{$\mathit{Ready(m_2)}$} & &  \\
		\hline
		& $\mathit{Ready(m_1)}$ &  & \textcolor{blue}{$\mathit{Ready(m_2)}$} & \textcolor{blue}{$\mathit{Ready(m_2)}$} \\
		\hline
		& blocked forever &  &  & $\mathit{Deliver(m_2)}$ \\ 
		\hline
	\end{tabular}
	\caption{Non-termination for Bracha protocol with blocking sets}
	\label{table:bracha-blocking}
\end{wraptable}

Quorum subsumption is inspired by and weakens the notion of quorum sharing \cite{losa2019stellar}.
%
Quorum sharing 
requires the above subsumption property for all quorums.
Thus, many quorum systems including Ripple and Stellar do not satisfy it
(unless Byzantine processes do not lie about their slices \cite{losa2019stellar}.)
%
%
They can maintain the subsumption property only for quorums of a well-behaved subset of processes.
In particular, no requirement can be made for quorums of Byzantine processes.
Therefore, we define the weaker notion of quorum subsumption for a subset of quorums,
and later show that it is sufficient to implement broadcast and consensus.



In order to make progress,
protocols (such as Bracha's Byzantine reliable broadcast \cite{bracha1985asynchronous}) 
require the members of a quorum to be able to communicate with at least one of their own quorums,
or communicate with a subset of processes that contains at least one well-behaved process.
Let us see intuitively how quorum subsumption can support liveness properties.
Consider a quorum system $𝓠$ for processes $𝓟 = \{1, 2, 3, 4\}$ 
where the Byzantine processes are $\{2\}$, 
and 
$𝓠(1) = \{\{1, 3, 4\}\}$, 
$𝓠(3) = \{\{1, 2, 3\}\}$, and 
$𝓠(4) = \{\{2, 3, 4\}\}$.
The quorum system $\Q$ has quorum intersection, and is weakly available for the set $\{ 1 \}$
since there is a well-behaved quorum $\{1, 3, 4\}$ for the process $1$.
%
%
In the classic Bracha protocol, 
the sender broadcasts $\mathit{Echo}(m)$,
a well-behaved broadcasts $\mathit{Echo}(m)$ when it receives it from the sender,
it broadcasts $\mathit{Ready}(m)$ 
after receiving $2f + 1$ $\mathit{Echo}(m)$ or $f + 1$ $\mathit{Ready}(m)$ messages,
and 
finally, delivers $m$ if it receives  $2f + 1$ $\mathit{Ready}(m)$ messages.
In Stellar \cite{lokhava2019fast} and follow-up works \cite{losa2019stellar, garcia2018federated, cachin2020asymmetric},
the check for
receiving $\mathit{Ready}(m)$ messages from $f + 1$ processes
is replaced with
receiving $\mathit{Ready}(m)$ messages from 
a blocking set of the current process. 
%
Let's consider the example execution presented in \autoref{table:bracha-blocking};
it gives an intuition of why the quorum system needs stronger conditions than weak availability.
Consider a Byzantine sender who sends $\mathit{BCast}(m₁)$ to process $\{ 1, 3, 4 \}$. 
Well-behaved process $\{1, 3, 4\}$ sends out $\mathit{Echo}(m₁)$ to each other.
We let process $1$ deliver $\mathit{Echo}(m₁)$ messages from process $1$, $3$, and $4$ first; it then sends out $\mathit{Ready}(m₁)$ messages.
We note that the two processes $3$, and $4$ cannot broadcast $\mathit{Ready}(m₁)$ since they have not received $\mathit{Echo}(m₁)$ from a quorum of their own.
Then the Byzantine process $2$ sends $\mathit{Ready}(m₂)$ messages to process $\{ 3, 4 \}$.
Since the set $\{ 2 \}$ is blocking for the quorums of both processes $3$ and $4$,
both send out $\mathit{Ready}(m₂)$ messages.
These broadcast protocols prevent a process that is ready for a value from getting ready for another value.
Therefore, although $\{3\}$ and $\{4\}$ are both blocking sets for the process $1$, 
it cannot become ready for $m₂$.
Process $1$ never receives enough 
Thus, $\mathit{Ready}$ messages for either $m₁$ or $m₂$ to deliver a message,
and is blocked forever.
%
%
If the quorum $\{1, 3, 4\}$ for $1$ had the quorum subsumption property,
then $3$ and $4$ 
could send out $\mathit{Ready}(m₁)$ messages,
and eventually $1$ would make progress.

\myparagraph{Complete Quorum}
We will later see that quorum availability and quorum subsumption are important together for liveness.
We succinctly combine the two properties into the notion of complete quorums.

\begin{definition}[Complete Quorum]\label{def:comple-quorum}
A quorum 
$q$ 
in a quorum system $\Q$
is a complete quorum
if all its members are well-behaved, and $\Q$ is quorum subsuming for $q$.
\end{definition}

In our previous running example \autoref{fig:running-example}, quorum $\{3, 4\}$ is a complete quorum: both of its members are well-behaved and $\Q$ is quorum subsuming for $\{3, 4\}$.

\begin{definition}[Strong Availability]
\label{def:available}
A quorum system $𝓠$ 
has strong availability for a subset of processes $P$
iff
every process in $P$ has at least one complete quorum. 
We call $P$ a \emph{strongly available set} for $𝓠$, and call a member of $P$ a \emph{strongly available process}.
We say that $\Q$ is strongly available if 
it is strongly available for a non-empty set.
\end{definition}

Intuitively, 
operations stay available at
a strongly available process 
since its complete quorum can perform operations on his behalf 
in the face of
Byzantine attacks.
%
%
%
In \autoref{fig:running-example}, 
$\Q$ is strongly available for $\{3, 4\}$.
In contrast, $\Q$ is only weakly available for process $1$,
since its quorum $\{1, 2, 3\}$ includes $2$ that is not well-behaved,
and its other quorum $\{1, 4\}$ is well-behaved but not a complete quorum.

By \autoref{lem:block-w-available},
every blocking set of
every strongly available process
contains at least one well-behaved process.

%% file: ProblemDefinitions.tex
\section{Protocol Implementation}
\label{sec:proto-impl}

In the subsequent sections, we will see that it is impossible to construct a protocol for Byzantine reliable broadcast and consensus in an HQS given only quorum intersection and quorum availability.
After that, we give a protocol for Byzantine reliable broadcast and consensus for an HQS that 
has quorum intersection and strong availability.
We first need a model of quorum-based protocols, 
and then the exact specifications of the distributed abstractions we aim to design protocols for.
In this section, we consider the former.

We consider a modular design for protocols.
A protocol is captured as a component that
accepts request events and issues response events.
A component uses other components as sub-components:
it issues requests to them and accepts responses from them.
A component stores a state and
defines handlers for
incoming requests from the parent component,
and incoming responses from children components.
Each handler gets the pre-state and the incoming event as input,
and outputs the post-state and outgoing events, 
either as responses to the parent or requests to the children components.
%
The outputs of a handler can be deterministically a function of its inputs, or randomized.

\begin{definition}[Determinism]
    \label{def:deterministic}
    A protocol
    is deterministic 
    iff
    the outputs of its handlers are a function of the inputs.
\end{definition}




\myparagraph{Quorum-based Protocols}
A large class of protocols are implemented based on quorum systems.
In order to state impossibility results for these protocols,
we capture the properties of quorum-based protocols \cite{losa2019stellar,lamport2006lower} as a few axioms. 
Our impossibility results concern protocols that adhere to the 
necessity, sufficiency, and locality axioms.


A process in a quorum-based protocol should process a request 
only if
it can communicate with at least one of its quorums.

\begin{axiom}[Necessity of Quorums \cite{losa2019stellar}]\label{def:quorum-based-safety} \ 
If a well-behaved process $p$ issues a response for a request 
then there must be a quorum $q$ of $p$ such that 
$p$ receives at least one message from each member of $q$.
\end{axiom}







In a quorum-based protocol,
a process only needs the participation of itself and members of one of its quorums
to deliver a message.

\begin{axiom}[Sufficiency of Quorums]
\label{def:sufficiency-of-quorums}
For every execution where a well-behaved process $p$ issues a response,
there exists an execution where only $p$ and  a quorum of $p$ take steps, 
and $p$ eventually issues the same response.    
\end{axiom}

We add a remark for Byzantine reliable broadcast (BRB) which has a designated sender process.
We will use a slight variant of the sufficiency axiom for BRB that states that
there exists an execution where only {∗the sender}, $p$ and a quorum of $p$ take steps.

A process's local state is only affected by 
the information that it receives from the members of it's quorums.

\begin{axiom}[Locality]
    \label{def:quorum-based-locality}
    The state of a well-behaved process changes upon receiving a message only if the sender is a member of one of its quorums.
\end{axiom}

For BRB,
we will use a slight variant of the locality axiom that
allows processes change state upon receiving messages from {∗the sender}
in addition to members of quorums.

%


\section{Protocol Specification}
\label{sec:protocol-spec}

We now define the specification of reliable broadcast and consensus for HQS.
The liveness properties are weaker than classical notions since
in an HQS, availability might be maintained only for a subset $P$ of well-behaved processes.


\myparagraph{Reliable Broadcast}
We now define the specification of the reliable broadcast abstraction.
The abstraction accepts a single broadcast request from a designated sender (either in the system or a process that is separate from the other processes in system), and issues delivery responses.

%
%
%
%

\begin{definition}[Specification of Reliable Broadcast] \ 
    \label{def:brb-spec}
    \label{def:agreement-rb}
    \label{def:validity-rb}
    \begin{itemize}
        \item 
        (Validity for a set of well-behaved processes $P$).
        If a well-behaved process $p$ broadcasts a message $m$, 
        then every process in $P$
        eventually delivers $m$.



        \item
        (Integrity).
        If a well-behaved process delivers a message $m$ from a well-behaved sender $p$, 
        then $m$ was previously broadcast by $p$.

        \item 
        (Totality for a set of well-behaved processes $P$).
        If a message is delivered by a well-behaved process,
        then every process in 
        $P$ eventually delivers a message.


        \item 
        (Consistency).
        No two well-behaved processes deliver different messages.

        \item
        (No duplication).
        Every well-behaved process delivers at most one message.        

    \end{itemize}
\end{definition}

%
We also consider a variant of reliable broadcast called federated voting.
Similar to reliable broadcast, 
the abstraction accepts a broadcast request from processes,
and issues delivery responses.
In contrast to reliable broadcast where there is a dedicated sender,
in federated voting, every process can broadcast a message.
The specification of federated voting is similar to that of reliable broadcast except for validity.
The messages that well-behaved processes broadcast may not be the same.
Therefore, the validity property provides guarantees only when the messages are the same or there is only one sender.
The validity property for a set of well-behaved processes $P$
guarantees that
if all well-behaved processes broadcast a message $m$, 
or
only one well-behaved process broadcasts a message $m$,
then every process in $P$ eventually delivers $m$.

%
%
%
%
%
%

\myparagraph{Consensus}
We now consider the specification of the consensus abstraction.
It accepts propose requests from processes in the system, and issues decision responses.

\begin{definition}[Specification of Consensus] \ 
    \label{def:consensus-spec}    
    \label{def:agreement-within-s}    
    \label{def:validity-within-s-C}
    \label{def:termination}
    \begin{itemize}
        \item 
%
         (Validity).
         If all processes are well-behaved, and some process decides a value, 
         then that value was proposed by some process.

        \item
        (Agreement).
        No two well-behaved processes decide differently.        
        

        \item
        (Termination for a set of well-behaved processes $P$).
        Every process in $P$ eventually decides.


        
    \end{itemize}
\end{definition}

\newpageprime

%% file: impossibility.tex
\section{Impossibility}


We now present the impossibility results for consensus and Byzantine Reliable Broadcast (BRB).
It is known that quorum intersection and quorum availability are necessary conditions 
\cite{losa2019stellar}
to implement consensus and BRB protocols.
In this section, we show that while these two conditions are necessary, they are not sufficient.

We consider
the information-theoretic settings (Fault axiom \cite{fischer1986easy}), where
byzantine processes have unlimited computational power,
and can show arbitrary behavior.
%
However,
processes communicate only over secure channels so that the recipient knows the identity of the sender.
A Byzantine process is unable to impersonate a well-behaved process.
%
This is similar to the classic unauthenticated Byzantine general problem \cite{lamport1982byzantine},
and
is necessary for open decentralized blockchains and HQS, where 
the trusted authorities including public key infrastructures may not be available.


The two proofs will take a similar approach.
First, we assume there does exist a protocol for our distributed abstraction that satisfies all the desired specifications.
We then present a quorum system $\Q$ 
and consider its executions
that have quorum intersection and availability
in the face of Byzantine attacks.
We then show through a series of indistinguishable executions
that the protocol cannot satisfy all the desired specifications, leading to a contradiction.
The high-level idea is that in the information-theoretic setting, 
a well-behaved process is not able to distinguish between an execution where the sender is Byzantine and sends misleading messages,
and an execution where the relaying process is Byzantine and forwards misleading messages.
For example, let $p_1, p_2$ and $p_3$ be three processes in the system.
When $p_3$ receives conflicting messages from $p_1$ through $p_2$, 
it does not know whether $p_1$ or $p_2$ is Byzantine.
This eventually leads to violation of the agreement or validity property of the abstraction.

We consider binary proposals for consensus, and binary values (from the sender) for reliable broadcast.
For the consensus abstraction, we succinctly present the values that processes propose as
as a vector of values that we call a configuration.
If the initial value of a process is $\bot$ in the configuration, that process is considered Byzantine.
Otherwise, the process is well-behaved. 
For example, a configuration $C = \langle0, 0, \bot \rangle$ denotes the first and second process proposing zero and the third process being Byzantine.

\subsection{Consensus}
\label{sec:consensus-imp}

We first consider consensus protocols in HQS.

\newpageprime


\begin{theorem}
    \label{thm:quorum-based-consensus}
    Quorum intersection and weak availability are not sufficient for
    deterministic quorum-based consensus protocols to provide 
    validity, agreement and termination for weakly available processes.
    
    
\end{theorem}


\begin{proof}

We suppose there is a quorum-based consensus protocol that guarantees 
validity, agreement, and termination for every quorum system $\Q$ with quorum intersection and weak availability, towards contradiction.
Consider a quorum system $\Q$ for processes $\P = \{a,b,c\}$ with the following quorums:
$\Q(a) = \{\{a,c\}\}$, $\Q(b) = \{\{a,b\}\}$, $\Q(c) = \{\{b,c\}\}$.

We make the following observations:
(1) if all processes are well-behaved, 
then $\Q$ has quorum intersection and weak availability for $\{ a, b, c \}$,
(2) if only process $a$ is Byzantine,
then $\Q$ preserves quorum intersection, and weak availability for $\{ c \}$,
(3) if only process $c$ is Byzantine,
then $\Q$ preserves quorum intersection, and weak availability for $\{ b \}$.
Going forward, we implicitly assume termination for weakly available processes.

Now consider the following four configurations 
as shown in \autoref{fig:indist-execs}:
$C_0 = \langle{0, 0, 0} \rangle$, 
$C_1 = \langle{1, 1, 1} \rangle$,  
$C_2 = \langle{0, 1, \bot} \rangle$, and
$C_3 = \langle{\bot, 1, 1} \rangle$.
The goal is now to show a series of executions over the configurations so that at least one property of the protocol is violated.

\begin{itemize}
\item 
We begin with execution $E_0$ (shown in red) with the initial configuration $C_0$.
All the messages between $a$ and $c$ are delivered.
By 
termination for weakly available processes
and
validity, process $a$ decides 0.
Additionally, by quorum sufficiency, $a$ can reach this decision with only processes $\{a,c\}$ taking steps.

\item Next, we have execution $E_1$ (shown in blue) with initial configuration $C_1$.
 All the messages between $b$ and $c$ are delivered.
Again,
by termination for weakly available processes
and
validity, process $c$ decides 1.
By quorum sufficiency, $c$ can reach this decision with only processes $\{b,c\}$ taking steps.

\item 
Next, we have execution $E_2$ as a sequence of $E₁$ and $E₀$,
with initial configuration $C_2$.
Suppose messages between well-behaved processes $a$ and $b$ are delayed.
Byzantine process $c$ first replays $E_1$ with process $b$, 
then replays $E_0$ with process $a$.
This cause process $a$ to decide $0$.
Now let Byzantine process $c$ stay silent, and messages between processes $a$ and $b$ be delivered.
By termination for $b$, agreement and quorum sufficiency, 
process $a$ makes $b$ decide $0$ as well (shown in green).


\item 
Lastly, we have execution $E_3$ with initial configuration $C_3$.
Suppose messages between $b$ to $c$ are delivered in the beginning.
We let processes $\{b,c\}$ replay $E_1$; thus, $c$ decides 1.
Then, Byzantine process $a$ sends messages to $b$ as if it were at the end of $E_2$.
In turn, $b$ decides 0.
Thus, agreement is violated as two well-behaved processes decided differently.


\end{itemize}
\end{proof}

\begin{wrapfigure}{R}{0.5\textwidth}
	\label{fig:impossibility}
	\centering
	\scalebox{0.9}{
	\begin{tikzpicture}
		\node[] (start) {};
		\node[right=1.5cm of start] (a) {$a$};
		\node[right=1.5cm of a](b){$b$};
		\node[right=1.5cm of b](c){$c$};
		\node[below=0.2cm of start] (C1) {$C_0$};
		\node[below=0.2cm of C1] (C2) {$C_1$};
		\node[below=0.2cm of C2] (C3) {$C_2$};
		\node[below=0.2cm of C3] (C4) {$C_3$};
		
		\node[right=1.4cm of C1] (C1a) {$0$};
		\node[right=1.5cm of C1a] (C1b) {$0$};
		\node[right=1.5cm of C1b] (C1c) {$0$};
		
		\node[right=1.4cm of C2] (C2a) {$1$};
		\node[right=1.5cm of C2a] (C2b) {$1$};
		\node[right=1.5cm of C2b] (C2c) {$1$};
		
		\node[right=1.4cm of C3] (C3a) {$0$};
		\node[right=1.5cm of C3a] (C3b) {$1$};
		\node[right=1.5cm of C3b] (C3c) {$\bot$};
		
		\node[right=1.4cm of C4] (C4a) {$\bot$};
		\node[right=1.5cm of C4a] (C4b) {$1$};
		\node[right=1.5cm of C4b] (C4c) {$1$};
		
		\draw[red] (1.7,-0.3) rectangle (2.2,-0.8);
		\draw[red] (5.5,-0.3) rectangle (6,-0.8);
		\node[right=0.5cm of C1c] (E1) {$E_0$};
		
		\draw[blue] (3.6,-1.1) rectangle (6,-1.6);
		\node[right=0.5cm of C2c] (E2) {$E_1$};
		
		\draw[red] (1.7,-1.8) rectangle (2.2,-2.3);
		\draw[red] (5.45,-1.75) rectangle (6.05,-2.35);	
		\draw[blue] (3.6,-1.8) rectangle (6,-2.3);
		\draw[green] (1.6,-1.75) rectangle (4.1,-2.35);
		\node[right=0.5cm of C3] (E3) {$E_2$};
		
		\draw[blue] (3.6,-2.6) rectangle (6,-3.1);
		\draw[green] (1.6,-2.55) rectangle (4.1,-3.15);
		\node[right=0.5cm of C4] (E4) {$E_3$};

	\end{tikzpicture}
	}
	\caption{Indistinguishable Executions}
	\label{fig:indist-execs}
\end{wrapfigure}

\vspace{-3ex}
\paragraph{Indistinguishably}
We provide some intuition for the proof construction.
Ultimately, the problem lies in process $b$ not being able to distinguish whether process $a$ or process $c$ is the Byzantine process.
More specifically, both $E_2$ and $E_3$ begin with execution $E_1$.
Since process $b$ cannot distinguish between the two executions, it does not know which value to decide.
If process $b$ believes $E_2$ is the actual execution, then $b$ should decide 0 to agree 
with the decision of well-behaved process $a$.
However, if $E_3$ is the actual execution, then agreement is violated as process $c$ decided 1.
Conversely, if process $b$ believes $E_3$ is the actual execution, 
then $b$ should decide 1 to agree with 
the decision of well-behaved process $c$.
Then, if $E_2$ is the actual execution, agreement is violated as the well-behaved process $a$ decided 0.

We note that this proof could not be constructed if
there was quorum subsumption.
For example, if the process $b$ adds the quorum $\{a, b, c\}$,
then $\Q$ will have quorum subsumption for the quorum $\{a, b, c\}$ of $b$.
However, then by quorum subsumption, there will be no Byzantine process,
and the executions $E₂$ and $E₃$ cannot be constructed.
If the process $a$ adds the quorum $\{a, b\}$,
then it will have quorum subsumption.
However, then the process $a$ cannot Byzantine process anymore,
and the executions $E_3$ cannot be constructed.
Similarly, if the process $b$ adds the quorum $\{b, c\}$,
the executions $E_2$ cannot be constructed.

\subsection{Byzantine Reliable Broadcast}
\label{sec:reliable-broad-imp}

Now, we prove the insufficiency of quorum intersection and quorum availability for
Byzantine reliable broadcast.

For the reliable broadcast abstraction, we represent the initial configuration as an array of values received by the processes from the sender.
The sender is a fixed and external process in the executions, and is only used to assign input values for processes in the system, which are captured as the initial configurations. 
The sender does not take steps in the executions, 
and processes are not able to distinguish executions based on the sender.

\begin{theorem}
   \label{thm:quorum-based-rb}
   Quorum intersection and weak availability are not sufficient for
   deterministic quorum-based reliable broadcast protocols to provide validity and totality for weakly available processes, and consistency.
\end{theorem}

\begin{proof}
The proof is similar to the proof for consensus.
In fact, we will reuse the construction.
There are differences between reliable broadcast and consensus specifications in 
(1) their validity properties, 
and 
(2) their totality and termination properties respectively.
The proof can be adjusted for these differences.
For reliable broadcast, we need a sender process $s$ who broadcasts a message.
In executions that we want a well-behaved process to deliver the message $m$,
we either
(1) keep the sender $s$ well-behaved and have it send $m$, and then apply validity,
or 
(2) have a process deliver $m$, then apply totality and consistency.
%
The initial configuration represents values received by each process from the sender.

Executions follow those in the previous proof.
Message delivery and delays mirror the previous executions.
In execution $E_0$ for configuration $C_0$, the well-behaved sender $s$ broadcasts 0, and messages between processes $a$ and $c$ are delivered.
By validity for weakly available processes, process $a$ delivers 0, and by quorum sufficiency, only processes $\{a,c\}$ need to take steps.
In execution $E_1$ for configuration $C_1$, the well-behaved sender $s$ broadcasts 1, and messages between processes $b$ and $c$ are delivered.
By validity for weakly available processes, and quorum sufficiency, process $c$ delivers 1, only with $\{b,c\}$ taking steps.
In configurations $C_2$ and $C_3$, the sender $s$ is Byzantine.
The messages between processes $a$ and $b$ are delayed in the beginning.
In execution $E_2$ for configuration $C_2$, 
the Byzantine sender $s$ and Byzantine process $c$ 
replay $E_1$ with process $b$, 
then replay $E_0$ with process $a$.
Then Byzantine process $c$ stays silent, and messages between processes $a$ and $b$ are delivered.
By totality for weakly available processes, since process $a$ delivers 
0, then process $b$ will also deliver a value.
By consistency, process $b$ delivers 0 as well.
In the last execution $E_3$ for configuration $C_3$,
we let the Byzantine process $a$ stay silent in the beginning, and processes $b$ and $c$ replay $E_1$. Thus, process $c$ delivers 1.
Afterwards, messages between process $b$ and $c$ are delayed, 
and the Byzantine process $a$ replays $E_2$.
%
Again, process $b$ cannot distinguish between the two executions $E_2$ and $E_3$.
Since process $a$ sends the exact same messages to process $b$ as the end of $E_2$, process $b$ will deliver 0.
Thus, consistency between $c$ and $b$ is violated.
\end{proof}

%% file: ProtocolsEric2.tex
\section{Protocols}
\label{sec:proto}

We just showed that quorum intersection and 
availability are not sufficient to implement our desired distributed abstractions.
Now, we show that 
quorum intersection and strong availability,
our newly introduced property
are sufficient to implement both Byzantine reliable broadcast and consensus.


\begin{figure}
\begin{algorithm}[H]
	\small
    \caption{Byzantine Reliable Broadcast (BRB)}
    \label{alg:brb}
    \DontPrintSemicolon
    \SetKwBlock{When}{when received}{end}
    \SetKwBlock{Upon}{upon}{end}
    \begin{multicols}{2}
    \Implements $\colon$ \ $\mathsf{Reliable Broadcast}$ \;
    \ \ \ $\request : \mathit{broadcast}(v)$ \;
    \ \ \ $\response : \mathit{deliver}(v)$
    
    \textbf{Vars:}  \;
    \ \ \ $Q$ \AComment{Minimal quorums of $\self$ \ \ \ }   
    \ \ \ $\F : \mathsf{Set}[𝓟]$ \AComment{The followers of $\self$ \ \ \ }
    \ \ \ $\mathit{echoed}, \mathit{readied}, \mathit{delivered} : \mathsf{Boolean} ←\false$ \;
    \ \ \ $E, R : V ↦ \mathsf{Set}[𝓟] ← ∅$ \; \AComment{Set of echoed and readied processes \ \ \ }
    
    \Uses $\colon$ \;
    \ \ \ $\apl : \mathsf{PointToPointLink}$ \;
    
    \Upon(\request \ \mbox{$\mathit{broadcast}(v)$ from sender} \label{alg:rb-request}) {
        $\apl \ \request \ \mathit{send}(p, \mathit{BCast}(v))$ for each $p ∈ \P$ \; \label{alg:rb-to-all}
    }
    
    \Upon($\apl \ \response \ \mathit{deliver}❪p', \mathit{BCast}❪v❫❫$ \label{alg:rb-received}){
      \If{$¬\mathit{echoed}$}{
        $\mathit{echoed} \leftarrow \true$ \;
        \textcolor{blue}{$\apl \ \request \ \mathit{send}(p, \mathit{Echo}(v))$ for each $p \in \F$ \;} \label{alg:rb-echo-send}
      }        
    }
            
    \Upon($\apl \ \response \ \mathit{deliver}❪p', \mathit{Echo}❪v❫❫$\label{alg:rb-echo-rec-q}){
      $E(v) ← E(v) ∪ \{ p' \}$ \;
      
      \If{$¬\mathit{readied}$ $∧$ \textcolor{blue}{$∃q ∈ Q. \ q ⊆ E(v)$}}{
        $\mathit{readied} \leftarrow \true$ \;
        \textcolor{blue}{$\apl \ \request \ \mathit{send}(p, \mathit{Ready}(v))$ for \;
        \Indp\nonl each $p \in \F$ \;} \label{alg:rb-ready-send}
      }
    }

    \Upon($\apl \ \response \ \mathit{deliver}❪p', \mathit{Ready}❪v❫❫$\label{alg:rb-ready-rec}){
      $R(v) ← R(v) ∪ \{ p' \}$ \;
      \If{$¬\mathit{readied}$ $∧$ \textcolor{blue}{$R$ is a blocking set of $\self$}\label{alg:rb-ready-rec-blocking}}{ 
        $\mathit{readied} ←\true$ \;
        \textcolor{blue}{$\apl \ \request \ \mathit{send}(p, \mathit{Ready}(v))$ for \;
        \Indp \nonl each $p \in \F$ \; } \label{alg:rb-ready-send2}
      }
      \If{$¬\mathit{delivered}$ $∧$ \textcolor{blue}{$∃q ∈ Q. \ q ⊆ R(v)$}\label{alg:rb-ready-rec-q}}{
         $\mathit{delivered} \leftarrow \true$ \;
         $\response \ \mathit{deliver}(v)$ \label{alg:rb-delivery-response}\;
      }
    }
%
\end{multicols}
\vspace{0.5em}
\end{algorithm}
\end{figure}

 
\subsection{Reliable Broadcast Protocol}
\label{sec:proto-reliable-broad}
In \autoref{alg:brb},
we adapt the Bracha protocol \cite{bracha1985asynchronous} to show that quorum intersection and strong availability together are sufficient for Byzantine reliable broadcast. 
The parts that are different from the classical protocol are highlighted in blue.

Each process stores the set of its individual minimal quorums $Q$,
and its set of followers $𝓕$.
It also stores the boolean flags 
$\mathit{echoed}$, $\mathit{readied}$, and $\mathit{delivered}$
which record actions the process has taken to avoid duplicate actions.
It further uses point-to-point links $\apl$ to each of its followers.
Upon receiving a request to broadcast a value $v$ (at \autoref{alg:rb-request}), 
the sender broadcasts the value $v$ to all processes (at \autoref{alg:rb-to-all}).
Upon receiving the message from the sender (at \autoref{alg:rb-received}), 
a well-behaved process echoes the message among its followers (at \autoref{alg:rb-echo-send})
only if it has not already $\mathit{echoed}$.
When a well-behaved process receives a quorum of consistent echo messages (at \autoref{alg:rb-echo-rec-q}),
it sends ready messages to all its followers (at \autoref{alg:rb-ready-send}). 
A well-behaved process can also send a ready message 
when it receives consistent ready messages from a blocking set (at \autoref{alg:rb-ready-rec-blocking}).
When a well-behaved process receives a quorum of consistent ready messages for $v$ (at \autoref{alg:rb-ready-rec-q}), it delivers $v$ 
(at \autoref{alg:rb-delivery-response}).
The implementation of the federated voting abstraction is similar.
The only difference is that there can be multiple senders
(at \autoref{alg:rb-request}).


We prove that this protocol implements Byzantine reliable broadcast when the quorum system satisfies 
quorum intersection, and strong availability.
We remember that 
strong availability requires both weak availability and quorum subsumption.
More precisely, it requires
a well-behaved quorum $q$ for a process $p$,
and 
quorum subsumption for $q$.


\begin{theorem}
\label{thm:quorum-based-rb-suffi}
Quorum intersection and strong availability are sufficient to implement Byzantine reliable broadcast.
\end{theorem}

This theorem follows from five lemmas in 
the appendix \cite{ourappendix}
that prove the protocol satisfies the specification of Byzantine reliable broadcast that we defined in \autoref{def:brb-spec}. 
Consider a quorum system with quorum intersection, and strong availability for $P$.
Here, we state and prove only the validity property.
%
\input{BRBValidity}


%

\input{consensus}

%% file: BRBValidity.tex
\begin{lemma}
    The BRB protocol guarantees validity for $P$.
\end{lemma}

\begin{proof}
    Consider a well-behaved sender that broadcasts a message $m$.
   We show that every process in $P$ eventually delivers $m$.
    By availability,
    every 
    process $p ∈ P$
    has a complete quorum $q$.
    Consider a process $p' \in q$.    
    By quorum subsumption, 
    $p'$ has a quorum $q' \subseteq q$.
   By availability, 
   all members of $q$ (including $q'$) are well-behaved.
%
   Thus, when they receive $m$ from the sender, 
   they all echo it to their followers.
%
    The processes in $q'$ have $p'$ as a follower.
    Thus, $p'$ receives consistent echo messages for $m$
    from one of its quorums $q'$.
    Thus, $p'$ sends out ready messages for $m$ to its followers.
    Thus, all processes in $q$ send out ready messages for $m$ to their followers.
   The processes in $q$ have $p$ as a follower.
   Therefore, $p$ receives a quorum of consistent ready messages for $m$ from one of its quorums $q$, 
   and delivers $m$.
%
%
%
\end{proof}

%% file: consensus.tex
\subsection{Byzantine Consensus Protocol}
\label{sec:proto-consensus}

In this section, we show that quorum intersection and strong availability
are sufficient to implement Byzantine consensus.
%
We first present the consensus protocol for heterogeneous quorum systems,
and then prove its correctness.

At a high level,
the protocol proceeds in rounds with assigned leaders for each.
Ballots that carry proposal values are totally ordered.
A leader tries to commit its own candidate ballot
only after aborting any lower ballot in the system.
Leaders use the federated voting abstraction 
(that we saw in \autoref{sec:protocol-spec})
to abort or commit ballots.
%
There may be multiple leaders or Byzantine leaders before GST,
and they may broadcast contradicting 
abort and commit
messages for the same ballot.
However, by the consistency property of federated voting, 
processes agree on aborting or committing ballots.

A ballot $b$ is a pair $〈r, v〉$
of
a round number $r$ and a proposed value $v$.
Ballots are totally ordered by first their round numbers, and then their values:
a ballot $\angled{r,v}$ is below another $\angled{r',v'}$,
written as
$〈r, v〉 < 〈r', v'〉$, if
$r < r'$ or $r = r' ∧ v < v'$.
Two ballots $b = 〈r, v〉$ and $b' = 〈r', v'〉$ are compatible, $b ∼ b'$,
if
they have the same value, \ie, $v = v'$;
otherwise, they are incompatible, $b \not∼ b'$.
We say that a ballot is below and incompatible with another, $b ⋦ b'$,
if $b < b'$ and $b \not∼ b'$.
For message passing communication, 
we assume batched network semantics (BNS), 
where 
messages issued in an event are sent as a batch,
and
the receiving process delivers and processes the batch of messages together.
(In particular, as we will see later in the correctness proofs, 
if prepare messages that are sent together are not processed together the validity property can be violated.)
The protocol is similar to SCP \cite{mazieres2015stellar,garcia2019deconstructing} in structure;
the important difference is that 
this protocol uses leaders 
\cite{losa2019stellar}
and guarantees termination.
%
Our protocol guarantees termination regardless of Byzantine processes.
On the other hand, 
the SCP protocol guarantees a liveness property called {∗non-blocking} 
which requires Byzantine processes to stop.
(More precisely,
if a process $p$ in the intact set 
\cite{mazieres2015stellar,garcia2018federated}
has not yet decided in some execution,
then for every continuation of that execution in which all the Byzantine processes stop, 
the process $p$ eventually decides.)

\begin{figure}
\begin{algorithm}[H]
\small
\caption{Byzantine Consensus 
	}
	\label{alg:bc}
	\DontPrintSemicolon
	\SetKwBlock{When}{when received}{end}
	\SetKwBlock{Upon}{upon}{end}
	\begin{multicols}{2}
	\Implements $\colon$ \ $\mathsf{Consensus}$ \;
	\ \ \ $\request : \mathit{propose}(v)$ \;
	\ \ \ $\response : \mathit{decice}(v)$
	
	\textbf{Vars:}  \;
	\ \ \ $\mathit{round}: \mathbb{N}^{+} \leftarrow 0$ \AComment{Current round number \ \ \ }
	\ \ \ $\mathit{candidate}, \mathit{prepared} : \langle \mathbb{N}^{+}, V \rangle ←\langle 0, \bot \rangle$ \;
	\ \ \ $\mathit{leader} : 𝓟 ←p₀$ \AComment{current leader  \ \ \ }
	
	\Uses $\colon$ \;
	\ \ \ $\mathit{fv} : B ↦ \mathsf{ByzantineReliableBroadcast}$ \;
	\ \ \ $\le : \mathsf{EventualLeaderElection}$ \;
	
	\Upon(\request \ \mbox{$\mathit{propose}(v)$} \label{algC:propose}) {
		$\mathit{candidate} \leftarrow \langle 1, v \rangle$ \label{algC:assign-proposal}\;
		\If{$\self = \mathit{leader}$}{			 
			$\mathit{fv}(b')$ \request $\, \mathit{broadcast}(\mathbb{A})$ for all $b' ⋦ \mathit{candidate}$\;  \label{algC:leader-abort}
		}
	}

	
	\Upon($\mathit{fv}❪\mathit{b'}❫ \ \response \ \mathit{deliver}❪p, \mathbb{A}❫$ for all $ b' ⋦ b$ where $\mathit{prepared} < b$ \label{algC:prepared}){
		$\mathit{prepared} \leftarrow b$ \label{algC:update-prepared}\;
			\If{$\self = \mathit{leader} ∧ \prepared = \candidate$\label{algC:prepare-candidate}}{
					$\mathit{fv}(\mathit{candidate})$ $\request$  $\mathit{broadcast}(\mathbb{C})$ \label{algC:leader-commit} 
         }
   }
	
	\Upon($\mathit{fv}\mbox{$(b)$} \ \response \ \mathit{deliver}❪p, \mathbb{C}❫$ where $b = \prepared \land p = \mathit{leader}$ \label{algC:commit-prepared}){
		$\response \ \mathit{decice}(b.v)$ \label{algC:decision-reponse}\;
	}

	\Upon($\mathit{timeout}$ triggered \label{algC:trigger-timer}){
		$\le \ \request \ \mathit{Complain}(\mathit{round})$ \label{algC:trigger-timer-prime}\;
	}
	
	\Upon($\le \ \response \ \mathit{new\text{-}leader}\mbox{($p$)}$\label{algC:new-leader}){
		$\mathit{leader} \leftarrow  p$ \;
		$\mathit{round} \leftarrow \mathit{round} + 1$ \label{algC:increase-round}\;
      \If{$\self = \mathit{leader}$\label{algC:if-delta}}{
         $\mbox{Delay for time } \Delta$ \label{algC:delta} \; 
      }
		start-timer($\mathit{round}$) \label{algC:reset-timer} \;
		\If{$\mathit{prepared} = \langle 0, \bot \rangle$}{
			$\mathit{candidate} \leftarrow \langle \mathit{round}, \mathit{candidate}.v\rangle$ \label{algC:update-candidate-default}\;
		}
		\Else{ 			
			$\mathit{candidate} \leftarrow \langle \mathit{round}, \mathit{prepared}.v\rangle$\label{algC:update-candidate} \;
		}
		\If{$\self = \mathit{leader}$\label{algC:self-new-leader}}{
         $\mathit{fv}(b')$ \request $\, \mathit{broadcast}(\mathbb{A})$ for all $b' ⋦ \mathit{candidate}$\; \label{algC:abort-new-round}
		}
	}
\end{multicols}
\vspace{0.5em}
\end{algorithm}
\end{figure}
Each process stores four local variables:
$\mathit{round}$ is the current round number,
$\candidate$ is the ballot that the process tries to commit,
$\prepared$ is the ballot that the process is safe to discard any ballots lower and incompatible with,
and
$\mathit{leader}$ is the current leader.
%
%
Each process uses an instance of federated voting for each ballot,
and
an eventual leader election module.
The latter issues $\mathit{new\text{-}leader}$ events, and
eventually elects a well-behaved process as the leader.
(Previous work \cite{losa2019stellar} presented a probabilistic leader election module.)


		Upon receiving a proposal request (at \autoref{algC:propose}),
		a well-behaved process initializes its candidate ballot to the pair of the first round and its own proposal (at \autoref{algC:assign-proposal}).
		If the current process $\self$ is the leader, it tries to prepare its $\candidate$ by broadcasting abort $\mathbb{A}$ messages for all ballots  with $\candidate$ (at \autoref{algC:leader-abort}).
		When a well-behaved process delivers $\mathbb{A}$ messages from the leader
		for all ballots below and incompatible with some ballot $b$, 
		and its current $\prepared$ ballot is below $b$ 
		(at \autoref{algC:prepared}), 
		it sets $\prepared$ to $b$ (at \autoref{algC:update-prepared}).
		If the current process $\self$ is the leader, 
		and the $\prepared$ ballot is equal to the $\candidate$ ballot,
		then it broadcasts a commit $\mathbb{C}$ message for its $\candidate$ ballot (at \autoref{algC:leader-commit}).
%
%
		When a 
		well-behaved 
		process delivers a $\mathbb{C}$ message for a ballot $b$ from the leader,
		and it has already prepared the same ballot (at \autoref{algC:commit-prepared}), 
		it decides the value of that ballot (at \autoref{algC:decision-reponse}).
	

To ensure liveness, a well-behaved process triggers a timeout 
if no value is decided after a predefined time elapses in each round.
The process then complains to the leader election module  (at \autoref{algC:trigger-timer-prime}).
When the leader election module issues a new leader (at \autoref{algC:new-leader}),
a well-behaved process 
updates its $\mathit{leader}$ variable,
and
increments the $\mathit{round}$ number (at \autoref{algC:increase-round}).
The leader itself then waits for a time $\Delta$ 
(at \autoref{algC:delta})
which we will further explain below.
The process also resets the timer with a doubled timeout
for the next round (at \autoref{algC:reset-timer}).
It then updates the $\candidate$ ballot:
if no value is prepared before, 
the $\candidate$ ballot is updated to the new round number and 
the value of 
the current $\candidate$ (at \autoref{algC:update-candidate-default});
otherwise, 
it is updated to the new round number and 
the value of the $\prepared$ ballot (at \autoref{algC:update-candidate}).
Then, the leader tries to prepare the $\candidate$ 
by aborting below and incompatible ballots
similar to the steps above
(at \autoref{algC:abort-new-round}).

\begin{wrapfigure}{R}{0.6\textwidth}
	\centering	
	\scalebox{.8}{
		\begin{tikzpicture}[font=\sffamily,>=stealth',thick,
			commentl/.style={text width=3cm, align=right},
			commentr/.style={commentl, align=left},]
			\node[] (l1) {\small \textcolor{red}{$l_1$}};
			\node[below=0.5cm of l1] (l2) {\small $l_2$};
			\node[below=0.5cm of l2] (p3) {\small $p_3$};
			\node[below=0.5cm of p3] (p4) {\small $p_4$};
			\node[right=9cm of l1](l1_end){};
			\node[right=9cm of l2](l2_end){};
			\node[right=9cm of p3](p3_end){};
			\node[right=9cm of p4](p4_end){};
			
			\node[label={[yshift=-0.8cm, xshift = 0.2cm] \small $\angled{1, 3}$}, right=0.1cm of l2](prepare_p1){};
			\node[label={[yshift=-0.8cm, xshift=0.2cm] \small $\angled{1, 4}$}, right=0.1cm of p3](prepare_p1){};
			\node[label={[yshift=-0.8cm, xshift=0.2cm] \small $\angled{1, 4}$}, right=0.1cm of p4](prepare_p1){};
			\node[right=0.3cm of l1](l1_commit_send){};
			\node[right=1.8 cm of p3](p3_commit_recv){};
			
			\node[right=3.8cm of l1, circle,fill,inner sep=1.5pt](l1_new_leader){};
			\node[right=3.9cm of l2, circle,fill,inner sep=1.5pt](l2_new_leader){};
			\node[right=4.1cm of p3, circle,fill,inner sep=1.5pt](p3_new_leader){};
			\node[right=3.8cm of p4, circle,fill,inner sep=1.5pt](p4_new_leader){};
			\node[right=7.0cm of l1, circle,fill,inner sep=1.5pt](l1_new_leader2){};
			\node[right=7.3cm of l2, circle,fill,inner sep=1.5pt](l2_new_leader2){};
			\node[right=7.1cm of p3, circle,fill,inner sep=1.5pt](p3_new_leader2){};
			\node[right=7.8cm of p4, circle,fill,inner sep=1.5pt](p4_new_leader2){};
			\node[right=9.5cm of l2](l2_new_leader3){};
			\node[right=9.5cm of p3](p3_new_leader3){};
			\node[right=9.5cm of p4](p4_new_leader3){};
			\node[label={[yshift=-0.8cm, xshift = -0.7cm] \small $\angled{1, 5}$}, right=0.1cm of l2_new_leader3](prepare_p1){};
			\node[label={[yshift=-0.8cm, xshift=-0.7cm] \small $\angled{1, 4}$}, right=0.1cm of p3_new_leader3](prepare_p1){};
			\node[label={[yshift=-0.8cm, xshift=-0.7cm] \small $\angled{1, 5}$}, right=0.1cm of p4_new_leader3](prepare_p1){};
			
			\node[right = 4.5cm of l1](l1_recv_abort){};
			\node[right = 5cm of p3](p3_recv_abort){};
			\node[right = 5.3cm of p4](p4_recv_abort){};
			\node[right = 2.8cm of l1](l1_recv_commit){};
			\node[right = 2.8cm of l2](l2_recv_commit){};
			\node[right = 2.8cm of p4](p4_recv_commit){};
			\node[right = 4.8cm of l2](l2_echo_send){};
			\node[right = 5.8cm of l1](l1_recv_echo_abort){};
			\node[right = 5.6cm of p3](p3_recv_echo_abort){};
			\node[right = 6cm of p4](p4_recv_echo_abort){};
			\node[right = 8.5cm of l2](l2_recv_echo_abort2){};
			\node[right = 8.5cm of p3](p3_recv_echo_abort2){};
			\node[right = 8.5cm of p4](p4_recv_echo_abort2){};
			
			\node[right = 6.8cm of l2](l2_recv_echo){};
			
			\draw[->] (l1_commit_send) -- (p3_commit_recv) node[pos=.4, above, sloped] {\scriptsize $\mathit{fv(b) \ BCast(\mathbb{C})}$};
			\draw[->] (p3_commit_recv) -- (l1_recv_commit) node[pos=.4, above, sloped] {};
			\draw[->] (p3_commit_recv) -- (l2_recv_commit) node[pos=.4, above, sloped] {};
			\draw[->] (p3_commit_recv) -- (p4_recv_commit) node[pos=.4, below, sloped] {\scriptsize $\mathit{fv(b) \ Echo(\mathbb{C})}$};
			\draw[->] (l2_new_leader) -- (l1_recv_abort) node[pos=.4, above, sloped] {};
			\draw[->] (l2_new_leader) -- (p3_recv_abort) node[pos=.4, above, sloped] {};
			\draw[->] (l2_new_leader) -- (p4_recv_abort) node[pos=.4, below, sloped] {\scriptsize $\mathit{fv(b) \ BCast(\mathbb{A})}$};
			\draw[->] (p4_recv_abort) -- (l2_recv_echo) node[pos=.4, above, sloped] {\scriptsize $\mathit{fv(b) \ Echo(\mathbb{A})}$};
			\draw[->] (l2_echo_send) -- (l1_recv_echo_abort) node[pos=.4, above, sloped] {\scriptsize $\mathit{fv(b) \ Echo(\mathbb{A})}$};
			\draw[->] (l2_echo_send) -- (p3_recv_echo_abort) node[pos=.4, above, sloped] {};
			\draw[->] (l2_echo_send) -- (p4_recv_echo_abort) node[pos=.4, above, sloped] {};
			\draw[->] (l1_new_leader2) -- (l2_recv_echo_abort2) node[pos=.4, above, sloped] {\scriptsize $\mathit{fv(b) \ Echo(\mathbb{A})}$};
			\draw[->] (l1_new_leader2) -- (p4_recv_echo_abort2) node[pos=.4, above, sloped] {};
			
			\draw[dashed, blue] (4,0.5) -- (4,-3.5);
			\draw[dashed, blue] (7.25,0.5) -- (7.25,-3.5);
			\node[below=1cm of p3_commit_recv] (l1_leader1) {\small \textcolor{red}{$l_1$}};
			\node[below=2cm of l2_echo_send] (l2_leader2) {\small $l_2$};
			\node[below=0.1cm of p4_new_leader2] (l1_leader3) {\small \textcolor{red}{$l_1$}};
			
			\draw[thick, shorten >=-1cm] (l1) -- (l1_end);
			\draw[thick, shorten >=-1cm] (l2) -- (l2_end);
			\draw[thick, shorten >=-1cm] (p3) -- (p3_end);
			\draw[thick, shorten >=-1cm] (p4) -- (p4_end);
		\end{tikzpicture}
		}
	\caption{Last Minute Attack. $b = \angled{1, 4}$. The $\candidate$ of well-behaved leader $l_2$ is $b' = \angled{2, 3}$. The votes $\mathbb{C}$ and $\mathbb{A}$ are abbreviated as $C$ and $A$. The new leader events are triggered at the black dots at each process. Prepared ballots are shown below the time line for each process.}
	\label{fig:proof-fig}
\end{wrapfigure}

Let us now explain why delay $Δ$ is needed for termination.
Without this delay, a Byzantine leader can perform a last minute attack
that we illustrate in \autoref{fig:proof-fig}.
Consider that we have four processes, one of them is Byzantine,
and any set of three processes is a quorum.
Let the Byzantine process be the leader $l₁$, 
and let the ballot $b$ be prepared.
The leader $l₁$ sends a commit for ballot $b$ to one well-behaved process $p₃$.
Then, $p₃$ 
echos commit for $b$.
Then,  the timeout for $l₁$ happens, and 
the next well-behaved leader $l₂$ comes up.
Without the delay, $l₂$ may have not prepared $b$ yet (although other well-behaved processes $p₃$ and $p₄$ prepared it).
Therefore,
the ballot $b'$ that $l₂$ updates its candidate to (at \autoref{algC:update-candidate})
is not $b$, and may not be compatible with $b$.
%
In order to prepare $b'$,
the leader $l₂$ tries to abort $b$ 
(at \autoref{algC:abort-new-round})
but $b$ cannot be aborted:
in order to abort $b$, a quorum of processes should echo it.
However, the well-behaved process $p₃$ has already echoed commit, 
and if the Byzantine process $l₁$ remains silent,
the remaining two well-behaved processes $l₂$ and $p₄$ are not a quorum,
and cannot abort $b$.
Therefore, $l₂$ cannot succeed, and the timeout is triggered.
Further, if the next leader is the Byzantine process $l₁$ again,
it can repeat the above scenario:
it can abort $b$ to prepare a higher ballot $b₂$,
and make a well-behaved process echo commit for $b₂$,
before passing the leadership.
The attack can continue infinitely, and delay termination.
If the delay $Δ$ is larger than the bounded communication delay after GST,
it makes the leader $l₂$ observe the highest prepared ballot $b$,
and adopt its value as the value of its candidate $b₂'$ (at \autoref{algC:update-candidate}).
When it tries to commit $b₂'$, 
since it is compatible with $b$, it does not need abort it.
Therefore, it can prepare and commit $b₂'$, and decide.
We also note that instead of the delay $Δ$,
the above attack can be avoided
if the leader election can provide two successive well-behaved leaders.


\begin{theorem}
   \label{thm:quorum-based-c-suffi}
   Quorum intersection and strong availability are sufficient to implement consensus.
\end{theorem}


This theorem follows from three lemmas in 
the appendix \cite{ourappendix}
that prove that the protocol satisfies the specification of Byzantine consensus
that we defined in \autoref{def:consensus-spec}. 
%
An example execution of the protocol is described 
in the appendix \cite{ourappendix}.

%% file: related_work.tex
\section{Related Works} 

\textbf{∗Quorum Systems with Heterogeneous Trust. \ }
%
%
Ripple \cite{schwartz2014ripple} and
Cobalt \cite{macbrough2018cobalt}
pioneered decentralized trust.
They let each node specify a list, called the unique node list (UNL), of processes that it trusts.
However, 
they do not consider quorum availability or subsumption.

Stellar \cite{mazieres2015stellar,lokhava2019fast}
presents federated Byzantine quorum systems (FBQS)
\cite{garcia2018federated,garcia2019deconstructing}
where 
quorums are iteratively calculated from quorums slices.
Stellar also presents a federated voting and consensus protocol.
In comparison,
the assumptions of the protocols presented in this paper are weaker, 
and their guarantees are stronger.
The stellar consensus protocol (SCP) guarantees termination when Byzantine processes stop.
In contrast, the consensus protocol in this paper guarantees termination regardless of Byzantine processes.
Further, abstract SCP \cite{garcia2018federated}
provides agreement only for intact processes.
The intact set for an FBQS is a subset of processes that have strong availability.
On the other hand,
the consensus protocol in this paper 
provides agreement for all well-behaved processes.
%
In FBQS, the intersections of quorums should have a process in the intact set;
however, in HQS, they only need to have a well-behaved process.
The validity and totality properties for the reliable broadcast for FBQS are restricted to
the intact set.
On the other hand,
%
the reliable broadcast protocol in this paper provides
%
totality for all processes that have weak availability,
and
validity for all processes that have strong availability.

Personal Byzantine quorum systems (PBQS) \cite{losa2019stellar} 
capture the quorum systems that FBQSs derive form slices,
and
propose a 
responsiveness
consensus protocol \cite{yin2019hotstuff,abraham2020information,pass2018thunderella,alistarh2019extension}.
%
They define a notion called quorum sharing
which requires quorum subsumption for every quorum.
Stellar quorums have quorum sharing
if and only if processes do not lie about their slices.
(The appendix \cite{ourappendix} presents examples.)
%
In this paper, 
we relax quorum sharing to quorum subsumption,
and capture quorums that FBQSs derive even when Byzantine quorums lie about their slices,
and
show that even if a quorum system does not satisfy quorum sharing,
safety can be maintained for all processes,
and liveness can be maintained for the set of strongly available processes.



Asymmetric Byzantine quorum systems (ABQS)
\cite{cachin2020asymmetric,cachin2020symmetric,alpos2021trust}
allow each process to define a subjective dissemination quorum system (DQS),
in a globally known system.
The followup model \cite{cachin2022quorum}
lets each process
specify a subjective DQS 
for processes that it knows,
transitively relying on the assumptions of other processes.
In contrast, HQS
lets each process specify its own set of quorums
without knowing the quorums of other processes.
Further, it
does not require the specification of a set of possible Byzantine sets.
%
Further, there are systems where a strongly available set (from HQS) exists but no guild set (from ABQS) exists. 
(The appendix \cite{ourappendix} presents examples.)
Therefore, HQS can provide safety and liveness for those executions but ABQS cannot.
%
%
ABQS presents shared memory and broadcast protocols,
and 
further, 
rules to compose two ABQSs.
%
On the other hand,
this paper 
proves impossibility results, and presents protocols for reliable broadcast and consensus abstractions.
%
HQS provides strictly stronger guarantees with weaker assumptions. 
In ABQS, the 
properties of reliable broadcast 
are stated for wise processes and the guild.
However, 
this paper states these four properties 
for well-behaved processes and the strongly available set.
Well-behaved processes are a superset of wise processes,
and as noted above, in certain executions, 
the strongly available set is a superset of the guild.


Flexible BFT \cite{malkhi2019flexible} 
allows different failure thresholds between learners.
Heterogeneous Paxos \cite{sheff2021heterogeneous,sheff2014distributed} further generalizes the separation between learners and acceptors with different trust assumptions;
it specifies quorums as sets rather than number of processes.
These two projects
introduce
a consensus protocol that
guarantees safety or liveness
for learners with correct trust assumptions.
%
However, they require the knowledge of all processes in the system.
In contrast, HQS only requires partial knowledge of the system, 
and captures the properties of quorum systems
where reliable broadcast and consensus protocols are impossible or possible.
Multi-threshold reliable broadcast and consensus \cite{hirt2020multi}
and MT-BFT \cite{momose2021multi}
elaborate Bracha \cite{bracha1985asynchronous}
to have different fault thresholds for different properties, and different synchrony assumptions.
However, they have cardinality-based or uniform quorums across processes.
In contrast, HQS 
supports
heterogeneous quorums.

%
%
K-consistent reliable broadcast (K-CRB) \cite{bezerra2022relaxed} introduces a relaxed reliable broadcast abstraction where the correct processes can define their own quorum systems.
Given a quorum system, it focuses on
delivering the smallest number $k$ of different values.
%
In contrast, we propose the weakest condition 
to solve classical reliable broadcast and consensus.
Moreover, K-CRB's relaxed liveness guarantee (accountability) requires public key infrastructure.
In contrast, all the results
in this paper are 
for information-theoretic setting.

Our consensus protocol uses eventual leader election.
%
Several other works 
present
view synchronization and eventual leader election 
for Byzantine replicated systems
\cite{bravo2022making,bravo2022liveness},
and dynamic networks
\cite{mostefaoui2005static,ingram2009asynchronous}.
It is interesting to see 
if their leader election modules can be generalized to the heterogeneous setting,
and support 
responsiveness 
\cite{yin2019hotstuff,attiya1994bounds}
for our consensus protocol.

\textbf{∗Impossibility Results.\ }
There are two categories of assumptions about the computational power of Byzantine processes.
In the information-theoretic setting, Byzantine process have unlimited computational resources.
While in the computational setting, Byzantine processes can not break a polynomial-time bound \cite{garay2020sok}.
In this work, our impossibility results for reliable broadcast and consensus fall in the information-theoretic category.
Whether the same results 
hold 
in the computational setting is an interesting open question. 

FLP \cite{fischer1985impossibility} proved that consensus is not solvable in asynchronous networks even with one crash failure.
Many following works \cite{goren2020probabilistic,delporte2004weakest,aguilera2006consensus,fischer1986easy,lamport1982byzantine,borcherding1996levels} considered solvability, and 
necessary and sufficient conditions for consensus and reliable broadcast to tolerate $f$ Byzantine failures in partially synchronous networks.
The number of processes should be more than $3f$ and the connectivity of the communication graph should be more than $2f$.
However, these results apply for cardinality-based quorums, which is a special instance of HQS.
We generalize the reliable broadcast and consensus abstractions
to HQS which supports non-uniform quorums,
and prove impossibility results for them.

%% file: conclusion.tex
\section{Conclusion}

This paper presented
a general model of 
heterogeneous quorum systems
where each process defines its own set of quorums,
and
captured their
properties.
Through indistinguishably arguments, 
it proved that 
no deterministic quorum-based protocol
can implement the consensus and Byzantine reliable broadcast abstractions
on a heterogeneous quorum system that provides only
quorum intersection and availability.
It introduced the quorum subsumption property,
and showed that the three conditions together are sufficient to implement the two abstractions.
It 
presented
Byzantine 
broadcast and consensus protocols
for heterogeneous quorum systems,
and proved their correctness
when the underlying quorum system maintain the three properties.

%% file: Doc.bbl
\begin{thebibliography}{10}

\bibitem{abraham2020information}
Ittai Abraham and Gilad Stern.
\newblock Information theoretic hotstuff.
\newblock {\em arXiv preprint arXiv:2009.12828}, 2020.

\bibitem{aguilera2006consensus}
Marcos~Kawazoe Aguilera, Carole Delporte-Gallet, Hugues Fauconnier, and Sam
  Toueg.
\newblock Consensus with byzantine failures and little system synchrony.
\newblock In {\em International Conference on Dependable Systems and Networks
  (DSN'06)}, pages 147--155. IEEE, 2006.

\bibitem{alistarh2019extension}
Dan Alistarh, James Aspnes, Faith Ellen, Rati Gelashvili, and Leqi Zhu.
\newblock Why extension-based proofs fail.
\newblock In {\em Proceedings of the 51st Annual ACM SIGACT Symposium on Theory
  of Computing}, pages 986--996, 2019.

\bibitem{alpos2021trust}
Orestis Alpos, Christian Cachin, and Luca Zanolini.
\newblock How to trust strangers: Composition of byzantine quorum systems.
\newblock In {\em 2021 40th International Symposium on Reliable Distributed
  Systems (SRDS)}, pages 120--131. IEEE, 2021.

\bibitem{attiya1994bounds}
Hagit Attiya, Cynthia Dwork, Nancy Lynch, and Larry Stockmeyer.
\newblock Bounds on the time to reach agreement in the presence of timing
  uncertainty.
\newblock {\em Journal of the ACM (JACM)}, 41(1):122--152, 1994.

\bibitem{baudet2019state}
Mathieu Baudet, Avery Ching, Andrey Chursin, George Danezis, Francois Garillot,
  Zekun Li, Dahlia Malkhi, Oded Naor, Dmitri Perelman, and Alberto Sonnino.
\newblock State machine replication in the libra blockchain.
\newblock {\em The Libra Assn., Tech. Rep}, 7, 2019.

\bibitem{bezerra2022relaxed}
Jo{\~a}o~Paulo Bezerra, Petr Kuznetsov, and Alice Koroleva.
\newblock Relaxed reliable broadcast for decentralized trust.
\newblock In {\em Networked Systems: 10th International Conference, NETYS 2022,
  Virtual Event, May 17--19, 2022, Proceedings}, pages 104--118. Springer,
  2022.

\bibitem{borcherding1996levels}
Malte Borcherding.
\newblock Levels of authentication in distributed agreement.
\newblock In {\em International Workshop on Distributed Algorithms}, pages
  40--55. Springer, 1996.

\bibitem{bracha1985asynchronous}
Gabriel Bracha and Sam Toueg.
\newblock Asynchronous consensus and broadcast protocols.
\newblock {\em Journal of the ACM (JACM)}, 32(4):824--840, 1985.

\bibitem{bravo2022liveness}
Manuel Bravo, Gregory Chockler, and Alexey Gotsman.
\newblock Liveness and latency of byzantine state-machine replication.
\newblock In {\em 36th International Symposium on Distributed Computing (DISC
  2022)}. Schloss Dagstuhl-Leibniz-Zentrum f{\"u}r Informatik, 2022.

\bibitem{bravo2022making}
Manuel Bravo, Gregory Chockler, and Alexey Gotsman.
\newblock Making byzantine consensus live.
\newblock {\em Distributed Computing}, 35(6):503--532, 2022.

\bibitem{buchman2016tendermint}
Ethan Buchman.
\newblock {\em Tendermint: Byzantine fault tolerance in the age of
  blockchains}.
\newblock PhD thesis, University of Guelph, 2016.

\bibitem{buchman2022revisiting}
Ethan Buchman, Rachid Guerraoui, Jovan Komatovic, Zarko Milosevic,
  Dragos-Adrian Seredinschi, and Josef Widder.
\newblock Revisiting tendermint: Design tradeoffs, accountability, and
  practical use.
\newblock In {\em 2022 52nd Annual IEEE/IFIP International Conference on
  Dependable Systems and Networks-Supplemental Volume (DSN-S)}, pages 11--14.
  IEEE, 2022.

\bibitem{cachin2022quorum}
Christian Cachin, Giuliano Losa, and Luca Zanolini.
\newblock Quorum systems in permissionless network.
\newblock {\em arXiv preprint arXiv:2211.05630}, 2022.

\bibitem{cachin2020asymmetric}
Christian Cachin and Bj{\"o}rn Tackmann.
\newblock Asymmetric distributed trust.
\newblock In {\em 23rd International Conference on Principles of Distributed
  Systems (OPODIS 2019)}. Schloss Dagstuhl-Leibniz-Zentrum f{\"u}r Informatik,
  2020.

\bibitem{cachin2020symmetric}
Christian Cachin and Luca Zanolini.
\newblock From symmetric to asymmetric asynchronous byzantine consensus.
\newblock {\em arXiv preprint arXiv:2005.08795}, 2020.

\bibitem{carr2022towards}
Harold Carr, Christa Jenkins, Mark Moir, Victor~Cacciari Miraldo, and Lisandra
  Silva.
\newblock Towards formal verification of hotstuff-based byzantine fault
  tolerant consensus in agda.
\newblock In {\em NASA Formal Methods: 14th International Symposium, NFM 2022,
  Pasadena, CA, USA, May 24--27, 2022, Proceedings}, pages 616--635. Springer,
  2022.

\bibitem{castro1999practical}
Miguel Castro, Barbara Liskov, et~al.
\newblock Practical byzantine fault tolerance.
\newblock In {\em OSDI}, volume~99, pages 173--186, 1999.

\bibitem{delporte2004weakest}
Carole Delporte-Gallet, Hugues Fauconnier, Rachid Guerraoui, Vassos Hadzilacos,
  Petr Kouznetsov, and Sam Toueg.
\newblock The weakest failure detectors to solve certain fundamental problems
  in distributed computing.
\newblock In {\em Proceedings of the twenty-third annual ACM symposium on
  Principles of distributed computing}, pages 338--346, 2004.

\bibitem{dwork1988consensus}
Cynthia Dwork, Nancy Lynch, and Larry Stockmeyer.
\newblock Consensus in the presence of partial synchrony.
\newblock {\em Journal of the ACM (JACM)}, 35(2):288--323, 1988.

\bibitem{fischer1986easy}
Michael~J Fischer, Nancy~A Lynch, and Michael Merritt.
\newblock Easy impossibility proofs for distributed consensus problems.
\newblock {\em Distributed Computing}, 1(1):26--39, 1986.

\bibitem{fischer1985impossibility}
Michael~J Fischer, Nancy~A Lynch, and Michael~S Paterson.
\newblock Impossibility of distributed consensus with one faulty process.
\newblock {\em Journal of the ACM (JACM)}, 32(2):374--382, 1985.

\bibitem{garay2020sok}
Juan Garay and Aggelos Kiayias.
\newblock Sok: A consensus taxonomy in the blockchain era.
\newblock In {\em Cryptographers’ track at the RSA conference}, pages
  284--318. Springer, 2020.

\bibitem{garcia2018federated}
{\'A}lvaro Garc{\'\i}a-P{\'e}rez and Alexey Gotsman.
\newblock Federated byzantine quorum systems.
\newblock In {\em 22nd International Conference on Principles of Distributed
  Systems (OPODIS 2018)}. Schloss Dagstuhl-Leibniz-Zentrum fuer Informatik,
  2018.

\bibitem{garcia2019deconstructing}
{\'A}lvaro Garc{\'\i}a-P{\'e}rez and Maria~A Schett.
\newblock Deconstructing stellar consensus (extended version).
\newblock {\em arXiv preprint arXiv:1911.05145}, 2019.

\bibitem{goren2020probabilistic}
Guy Goren, Yoram Moses, and Alexander Spiegelman.
\newblock Probabilistic indistinguishability and the quality of validity in
  byzantine agreement.
\newblock {\em arXiv preprint arXiv:2011.04719}, 2020.

\bibitem{hirt2020multi}
Martin Hirt, Ard Kastrati, and Chen-Da Liu-Zhang.
\newblock Multi-threshold asynchronous reliable broadcast and consensus.
\newblock {\em Cryptology ePrint Archive}, 2020.

\bibitem{ingram2009asynchronous}
Rebecca Ingram, Patrick Shields, Jennifer~E Walter, and Jennifer~L Welch.
\newblock An asynchronous leader election algorithm for dynamic networks.
\newblock In {\em 2009 IEEE International Symposium on Parallel \& Distributed
  Processing}, pages 1--12. IEEE, 2009.

\bibitem{lamport2006lower}
Leslie Lamport.
\newblock Lower bounds for asynchronous consensus.
\newblock {\em Distributed Computing}, 19:104--125, 2006.

\bibitem{lamport1982byzantine}
Leslie Lamport, Robert Shostak, and Marshall Pease.
\newblock The byzantine generals problem.
\newblock {\em ACM Transactions on Programming Languages and Systems}, pages
  382--401, 1982.

\bibitem{ourappendix}
Xiao Li, Eric Chan, and Mohsen Lesani.
\newblock Quorum subsumption for heterogeneous quorum systems. technical
  report.
\newblock In {\em International Symposium on Distributed Computing (DISC
  2023)}, 2023.

\bibitem{li2023open}
Xiao Li and Mohsen Lesani.
\newblock Open heterogeneous quorum systems, 2023.
\newblock \href {http://arxiv.org/abs/2304.02156} {\path{arXiv:2304.02156}}.

\bibitem{lokhava2019fast}
Marta Lokhava, Giuliano Losa, David Mazi{\`e}res, Graydon Hoare, Nicolas Barry,
  Eli Gafni, Jonathan Jove, Rafa{\l} Malinowsky, and Jed McCaleb.
\newblock Fast and secure global payments with stellar.
\newblock In {\em Proceedings of the 27th ACM Symposium on Operating Systems
  Principles}, pages 80--96, 2019.

\bibitem{losa2019stellar}
Giuliano Losa, Eli Gafni, and David Mazi{\`e}res.
\newblock Stellar consensus by instantiation.
\newblock In {\em 33rd International Symposium on Distributed Computing (DISC
  2019)}. Schloss Dagstuhl-Leibniz-Zentrum fuer Informatik, 2019.

\bibitem{macbrough2018cobalt}
Ethan MacBrough.
\newblock Cobalt: Bft governance in open networks.
\newblock {\em arXiv preprint arXiv:1802.07240}, 2018.

\bibitem{malkhi2019flexible}
Dahlia Malkhi, Kartik Nayak, and Ling Ren.
\newblock Flexible byzantine fault tolerance.
\newblock In {\em Proceedings of the 2019 ACM SIGSAC conference on computer and
  communications security}, pages 1041--1053, 2019.

\bibitem{malkhi1998byzantine}
Dahlia Malkhi and Michael Reiter.
\newblock Byzantine quorum systems.
\newblock {\em Distributed computing}, 11(4):203--213, 1998.

\bibitem{mazieres2015stellar}
David Mazieres.
\newblock The stellar consensus protocol: A federated model for internet-level
  consensus.
\newblock {\em Stellar Development Foundation}, 32:1--45, 2015.

\bibitem{miller2016honey}
Andrew Miller, Yu~Xia, Kyle Croman, Elaine Shi, and Dawn Song.
\newblock The honey badger of bft protocols.
\newblock In {\em Proceedings of the 2016 ACM SIGSAC conference on computer and
  communications security}, pages 31--42, 2016.

\bibitem{momose2021multi}
Atsuki Momose and Ling Ren.
\newblock Multi-threshold byzantine fault tolerance.
\newblock In {\em Proceedings of the 2021 ACM SIGSAC Conference on Computer and
  Communications Security}, pages 1686--1699, 2021.

\bibitem{mostefaoui2005static}
Achour Mostefaoui, Michel Raynal, Corentin Travers, Stacy Patterson, Divyakant
  Agrawal, and Amr~EL Abbadi.
\newblock From static distributed systems to dynamic systems.
\newblock In {\em 24th IEEE Symposium on Reliable Distributed Systems
  (SRDS'05)}, pages 109--118. IEEE, 2005.

\bibitem{nakamoto2008peer}
Satoshi Nakamoto.
\newblock Bitcoin: A peer-to-peer electronic cash system.
\newblock {\em White paper}, 2008.

\bibitem{pass2018thunderella}
Rafael Pass and Elaine Shi.
\newblock Thunderella: Blockchains with optimistic instant confirmation.
\newblock In {\em Advances in Cryptology--EUROCRYPT 2018: 37th Annual
  International Conference on the Theory and Applications of Cryptographic
  Techniques, Tel Aviv, Israel, April 29-May 3, 2018 Proceedings, Part II 37},
  pages 3--33. Springer, 2018.

\bibitem{schwartz2014ripple}
David Schwartz, Noah Youngs, and Arthur Britto.
\newblock The ripple protocol consensus algorithm.
\newblock {\em Ripple Labs Inc White Paper}, 5(8):151, 2014.

\bibitem{sheff2021heterogeneous}
Isaac Sheff, Xinwen Wang, Robbert van Renesse, and Andrew~C Myers.
\newblock Heterogeneous paxos.
\newblock In {\em OPODIS: International Conference on Principles of Distributed
  Systems}, number 2020 in OPODIS, 2021.

\bibitem{sheff2014distributed}
Isaac~C Sheff, Robbert van Renesse, and Andrew~C Myers.
\newblock Distributed protocols and heterogeneous trust: Technical report.
\newblock {\em arXiv preprint arXiv:1412.3136}, 2014.

\bibitem{veronese2011efficient}
Giuliana~Santos Veronese, Miguel Correia, Alysson~Neves Bessani, Lau~Cheuk
  Lung, and Paulo Verissimo.
\newblock Efficient byzantine fault-tolerance.
\newblock {\em IEEE Transactions on Computers}, 62(1):16--30, 2011.

\bibitem{yin2019hotstuff}
Maofan Yin, Dahlia Malkhi, Michael~K Reiter, Guy~Golan Gueta, and Ittai
  Abraham.
\newblock Hotstuff: {BFT} consensus with linearity and responsiveness.
\newblock In {\em Proceedings of the 2019 ACM Symposium on Principles of
  Distributed Computing}, pages 347--356, 2019.

\end{thebibliography}
